\documentclass[12pt,preprint]{mn2e}

\usepackage{graphicx}
\usepackage{wasysym}

\title[How robust is the lens-redshift test?]{How robust
are the constraints on cosmology and galaxy evolution from the
lens-redshift test?}

\author[Pedro R. Capelo and Priyamvada Natarajan]{Pedro
R. Capelo$^{1}$\thanks{e-mail: capelo@astro.yale.edu} and Priyamvada
Natarajan$^{1,2}$\thanks{e-mail: priya@astro.yale.edu}\\
$^{1}$Astronomy Department, Yale University, P.O. Box 208101, New
Haven, CT 06520-8101, U.S.A.\\ $^{2}$Department of Physics, Yale
University, P.O. Box 208101, New Haven, CT 06520-8101, U.S.A.}

\begin{document}

\date{}

\pagerange{\pageref{firstpage}--\pageref{lastpage}} \pubyear{2007}

\maketitle

\label{firstpage}

\begin{abstract}
The redshift distribution of galaxy lenses in known gravitational lens
systems provides a powerful test that can potentially discriminate
amongst cosmological models. However, applications of this elegant
test have been curtailed by two factors: our ignorance of how galaxies
evolve with redshift, and the absence of methods to deal with the
effect of incomplete information in lensing systems. In this paper, we
investigate both issues in detail. We explore how to extract the
properties of evolving galaxies, assuming that the cosmology is well
determined by other techniques. We propose a new nested Monte Carlo
method to quantify the effects of incomplete data. We apply the
lens-redshift test to an improved sample of seventy lens systems
derived from recent observations, primarily from the SDSS, SLACS and
the CLASS surveys. We find that the limiting factor in applying the
lens-redshift test derives from poor statistics, including incomplete
information samples, and biased sampling. Many lenses that uniformly
sample the underlying true image separation distribution will be
needed to use this test as a complementary method to measure the value
of the cosmological constant or the properties of evolving
galaxies. Planned future surveys by missions like the SNAP satellite
or LSST are likely to usher in a new era for strong lensing studies
that utilize this test. With expected catalogues of thousands of new
strong lenses, the lens-redshift test could offer a powerful tool to
probe cosmology as well as galaxy evolution.
\end{abstract}

\begin{keywords}
gravitational lensing -- galaxies: evolution -- galaxies: luminosity
function, mass function -- cosmological parameters -- cosmology:
observations -- cosmology: theory
\end{keywords}

\section{Introduction}

Gravitational lensing statistics have now been used to map the mass
distribution in galaxies (Blandford \& Narayan 1992; Narayan \&
Bartelmann 1996; Kochanek 2004) as well as to constrain cosmological
parameters (Cheng \& Krauss 1998; Chae 2003; Maoz 2005). Since the
discovery of the first multiply imaged quasar (Walsh, Carswell \&
Weymann 1979), well over a hundred such systems have now been
discovered in various wave-bands, ranging from the optical to the
radio. This progress is attributed to several dedicated on-going all
sky surveys like the CLASS (Myers et al. 2003), the SDSS (York et
al. 2000), the Two-Degree Field Galaxy Redshift Survey (2dFGRS;
Colless et al. 2001) and more recently the SLACS (Bolton et
al. 2006). Consequently, there has also been significant progress in
analytical and statistical studies of lenses. Many sophisticated
methods are now available to model strong lensing systems using both
parametric and non-parametric mass distributions (e.g. Rusin, Kochanek
\& Keeton 2003; Saha \& Williams 2003).

The concept of optical depth to lensing (ODTL) was proposed to study
strong lensing statistics by Turner, Ostriker \& Gott (hereafter TOG,
1984): they presented an analytic calculation of the lensing
probability of distant quasars by intervening galaxy lenses and the
role of selection effects therein. Since the lensing probability
depends on the comoving volume element, the ODTL test can be used to
constrain cosmological parameters by comparing the number of expected
lenses to the number of observed ones. Using the ODTL test, Kochanek
(hereafter K96, 1996) obtained limits on the cosmological constant
from the statistics of gravitational lenses using a number of
completed quasar surveys (e.g. Snapshot Survey, the ESO/Li\`{e}ge
survey, the NOT survey, the HST GTO survey, the FKS survey), lens
data, and a range of lens models. The formal limit obtained was
$\Omega_{\Lambda} < 0.66$ at 95\% confidence in flat cosmologies,
which included the statistical uncertainties in the number of lenses,
galaxies, quasars, and the parameters relating galaxy luminosities to
dynamical variables. This value is in contrast to what is now well
established by WMAP observations of the CMB (e.g. Spergel et al. 2006)
and high redshift SN Ia observations (Riess et al. 1998; Perlmutter et
al. 1999). These observations have in fact led to what is currently
referred to as a `cosmic concordance' model (Ostriker \& Steinhardt
1995) - the $\Lambda$-CDM model (with $h\simeq0.7$,
$\Omega_{\Lambda}\simeq0.7$, $\Omega_M\simeq0.3$ and
$\Omega_K\simeq0.0$) - as the most widely accepted description of the
Universe. It has been argued by K96 that their retrieved low value for
$\Omega_{\Lambda}$ could be due to dust obscuration in a large
fraction of lensing galaxies: however, a hundred times more dust is
needed to change the expected number of lenses by a factor of
two. Given this extreme value, dust is clearly not the dominant source
of systematic errors. By tabulating various sources of error and the
limitations imposed on the accuracy of the determination of
$\Omega_{\Lambda}$, K96 speculated that the assumptions on the
velocity dispersion function of lenses might be a significant source
of error. Reviewing previous estimates of the cosmological constant
derived from strong lensing statistics Maoz (2005) concludes that the
discrepancies might be due to possibly a lower lensing cross section
for ellipticals galaxies than assumed in the past.  Maoz (2005) argues
that the current agreement between recent model calculations and the
results of radio lens surveys may be fortuitous, and due to a
cancellation between the errors in the input parameters for the lens
population and the cosmology, as well as input parameters for the
source populations.
 
In the quest to determine the correct underlying cosmological model by
placing better and tighter constraints on $\Omega_{\Lambda}$, strong
gravitational lensing has not been the most reliable
technique. Systematic errors have plagued the lensing analysis,
leading to contradictory results for the derived values of the
cosmological constant in a flat Universe (see, for example, Maoz \&
Rix 1993, K96, Chae et al. 2002). These contradictory results were
primarily caused by: small number statistics due to the shortage of
observed lens systems; assumptions about the relationship between
luminosities and masses of galaxies; scatter in the empirical relation
between mass and light; and observational biases, mainly the
magnification bias\footnote{The magnification bias arises due to the
fact that intrinsically faint sources can appear in a flux-limited
survey by virtue of gravitational lensing thereby affecting the
statistics.}. An explicit relation between mass and light is required
for the lensing analysis in the absence of independent mass estimates
for the lensing galaxies. The luminosity of galaxies is converted into
a mass distribution (which is the relevant quantity to model lensing
effects) using a density profile, which is parametrized via the
velocity dispersion. Statistics of strong lenses and any cosmological
constraints thereby obtained depend on the assumed velocity dispersion
function (VDF) of galaxies.

Kochanek (hereafter K92, 1992), devised a test, the `lens-redshift
test', which circumvented the magnification bias since it does not
involve computing the total ODTL. This test relies on the computation
of the \textit{differential optical depth to lensing} with respect to
the angular critical radius $r$. The probability distributions of lens
redshifts $z$ with a given angular critical radius,
$[{(d\tau/dz)}/\tau](r)$, are evaluated. However, this quantity still
required knowledge of the VDF of lensing galaxies, which was inferred
(hence IVDF) by combining the Schechter luminosity function with an
empirical relation between luminosity and velocity dispersion, the
Faber-Jackson and the Tully-Fisher relations for early-type and
late-type galaxies, respectively. The lens-redshift test depends on
cosmological parameters as well as on galaxy evolution
parameters. Therefore, it can be used to constrain the former by
fixing the latter, or vice versa. With the assumption of no evolution,
K92 derived $\Omega_{\Lambda} \apprle 0.9$.

Ofek, Rix \& Maoz (hereafter ORM, 2003) revived the lens-redshift test
(K92). They applied it to a larger sample of lens systems than were
available to the K92 analysis, using the CLASS and SDSS surveys.
Their study also included a re-derivation and generalisation of the
lens-redshift test, which incorporated mass and number density
evolution of lens galaxies. They explicitly included the redshift
evolution of the characteristic velocity dispersion and evolution of
the number density of galaxies. The limit obtained by ORM for a flat
Universe, assuming no mass evolution of early-type galaxies between
$z$ = 0--1, was $\Omega_{\Lambda}<0.95$ at the 99\% confidence
limit. Turning things around, and fixing the cosmological model to
$\Omega_{\Lambda}=0.7$ and $\Omega_M=0.3$, they determined galaxy
evolution parameters and found $\hbox{d log}_{10} \sigma_{\star}
(z)/dz = -0.10_{-0.6}^{+0.6}$ and $\hbox{d log}_{10} n_{\star} (z)/dz
= +0.7_{-1.2}^{+1.4}$, where $\sigma_{\star}$ and $n_{\star}$ are the
characteristic velocity dispersion and number density of lensing
galaxies, respectively.

Mitchell et al. (hereafter MKFS, 2005) focused instead on the ODTL
test (TOG). In addition to using a larger sample than the one used by
K96, they included the evolution of the VDF in amplitude and shape,
based on theoretical galaxy formation models, and used the measured
velocity distribution function (MVDF) for early-types from the SDSS
(Sheth et al. 2003). MKFS found $\Omega_{\Lambda}$ = 0.74--0.78 for a
flat Universe prior and a limit $\Omega_{\Lambda}<0.86$ at the 95\%
confidence limit. Including the effects of galaxy evolution, they
found $\Omega_{\Lambda}$ = 0.72--0.78 and a limit
$\Omega_{\Lambda}<0.89$ at the 95\% confidence limit.

The consequence of using the MVDF versus the IVDF in the determination
of $\Omega_{\Lambda}$ is one of the key questions we address in this
work. The IVDF and MVDF differ at high luminosities/velocity
dispersions (see Fig.~A1 in the Appendix). The scatter of the
Faber-Jackson relation was a predominant source of uncertainty in the
previous studies (cf. ORM), leading to a systematic underestimation of
the number of objects with large velocity dispersions.

In this paper, we investigate the lens-redshift test in detail and
re-examine the uncertainties that limit its use as a powerful
discriminant between cosmological models, as well as its potential to
constrain galaxy evolution models. We apply this to a new enlarged
sample of lenses. This is done for the first time using the measured
velocity dispersion function from SDSS although we compare and
reproduce the results of ORM using the inferred velocity dispersion
function. In addition, we consider the effect of incomplete lensing
information on the retrieval of cosmological parameters with a new
nested Monte Carlo method.

The outline of the paper is as follows. In section 2 we define the
lens-redshift test and compare the use of the IVDF and the MVDF on the
determination of both $\Omega_{\Lambda}$ and galaxy evolution
parameters. In section 3 we describe the new expanded sample and, in
section 4, we present the results of the application of the
lens-redshift test to our sample. We present a new Monte Carlo method
to quantify the effect of incomplete lensing information in section 5,
by constructing realizations of several biased subsamples. We conclude
with a discussion of our results and their implication for future
observational surveys.

\section{The lens-redshift test formalism}

\subsection{Methodology using the inferred velocity dispersion function}

We follow the notation introduced by TOG and K92 in defining the
optical depth to lensing and the lens-redshift test, respectively. The
differential optical depth to lensing per unit redshift is the
differential probability $d\tau$ that a line of sight intersects a
lens at redshift $z$ in traversing the path $dz$ from a population of
lensing galaxies with comoving number density $n_L$. Mathematically,
for a source this is simply the ratio of the differential light travel
distance $cdt$ to its mean free path between successive encounters
with galaxies $1/n_LS$,

\begin{equation}
\frac{d\tau}{dz}=n_L(z) S \frac{r_H}{E(z)(1+z)},
\end{equation}

\noindent where the comoving number density of lensing galaxies is
given by $n_L = n_{\star}(1+z)^3$; $n_{\star}$ is the average number
density of lensing galaxies; $S$ is the cross section for multiple
imaging of a background point source; $r_H=c/H_0$ is the Hubble radius
and
$E(z)=(\Omega_M(1+z)^3+\Omega_K(1+z)^2+\Omega_{\Lambda})^\frac{1}{2}$.
The cross section for multiple imaging $S$ is given by $ S=\pi
r^2{D_L}^2.$ We initially assume $n_{\star}$, the characteristic
luminosity $L_{\star}$ and the characteristic velocity dispersion
$\sigma_{\star}$ of lensing galaxies to be constant with redshift,
although we will later relax this assumption and allow for time
evolution.

Our analysis is restricted to early-type and S0 galaxies as lenses and
it is assumed that they can be modelled as singular isothermal spheres
(SIS) \footnote{A SIS has a mass distribution given by
$\rho(r)=\sigma^2/2\pi \hbox{G} r^2$, where $\sigma$ is constant with
radius $r$. This density profile is a very good fit for elliptical and
S0 lensing galaxies. In fact non-singular isothermal spheres and
truncated isothermal spheres give very similar fits to lensing data
(e.g. Rusin et al. 2003).}. With the assumptions stated above, we can
write the angular critical radius $r$ as:

\begin{equation}
r=4\pi\left(\frac{\sigma}{c}\right)^2\frac{D_{LS}}{D_S}{f_E}^2,
\end{equation}

\noindent where $D_{LS}$ and $D_S$ are the angular diameter distances
between the lens and the source and between the observer and the
source, respectively, and $f_E$ is a parameter that takes into account
the difference between the velocity dispersion of the mass
distribution $\sigma_m$ and the observed stellar velocity dispersion
$\sigma$: $\sigma_m=f_E\sigma$. Modelling galaxies as singular
isothermal spheres, the characteristic central velocity dispersions
(which are typically unmeasured for most lenses), are drawn from the
VDF.

In this subsection, we relate the luminosity distribution to the
Faber-Jackson law and construct the IVDF. Using the Schechter function
fit to model the luminosity function of lensing galaxies,

\begin{equation}
\frac{dn}{dL}=\frac{n_{\star}}{L_{\star}}\left(\frac{L}{L_{\star}}\right)^{\alpha}
\hbox{exp}\left[-\frac{L}{L_{\star}}\right],
\end{equation}

\noindent and the Faber-Jackson relation,
${L}/{L_{\star}}=\left({\sigma}/{\sigma_{\star}}\right)^{\gamma}$, to
relate the luminosity to a velocity dispersion, combining these two
equations we derive the IVDF:

\begin{equation}
\frac{dn}{d\sigma}=\frac{n_{\star}}{\sigma_{\star}}
\left(\frac{\sigma}{\sigma_{\star}}\right)^{\gamma\alpha+\gamma-1}
\hbox{exp}\left[-\left(\frac{\sigma}{\sigma_{\star}}\right)^\gamma\right]\gamma.
\end{equation}

Combining the IVDF with equation (1), the differential optical depth can be written as\\[-8mm]

\begin{eqnarray}
\frac{d\tau}{dz\left(d\sigma/\sigma_{\star}\right)} & = &
\frac{(1+z)^2r_H}{E(z)} S n_{\star} \gamma 
\left(\frac{\sigma}{\sigma_{\star}}\right)^{\gamma \alpha + \gamma -1} \nonumber \\[-2mm]
& & \\[-2mm]
& \times &
\hbox{exp}\left[-\left(\frac{\sigma}{\sigma_{\star}}\right)^{\gamma}\right]. \nonumber
\end{eqnarray}

Defining $r_{\star} \equiv 4\pi(\sigma_{\star}/c)^2$ and using

\begin{equation}
\frac{\sigma}{\sigma_{\star}}=\left(\frac{r}{r_{\star}}\frac{1}{{f_E}^2}
\frac{D_S}{D_{LS}}\right)^{\frac{1}{2}},
\end{equation}

\noindent gives us the IVDF lens-redshift test equation,\\[-8mm]

\begin{eqnarray}
\frac{d\tau}{dz dr} & = &
\tau_{\star}\frac{\gamma}{2}\left(\frac{r}{r_{\star}}
\frac{1}{{f_E}^2} \frac{D_S}{D_{LS}}\right)^{\frac{\gamma}{2}(\alpha+1)} \nonumber \\[-2mm]
& & \\[-2mm]
& \times &
\hbox{exp}\left[-\left(\frac{r}{r_{\star}} \frac{1}{{f_E}^2}
  \frac{D_S}{D_{LS}}
\right)^{\frac{\gamma}{2}}\right]
\frac{{D_L}^2(1+z)^2}{{r_H}^2E(z)}\frac{r}{{r_{\star}}^2}, \nonumber
\end{eqnarray}

\noindent where $r_{\star}$ and $\tau_{\star} \equiv
16{\pi}^3n_{\star}{r_H}^3(\sigma_{\star}/c)^4$ are constants.

Incidentally, we note that this is slightly different in form from
K92, as we compute ${d\tau/dz dr}$, whereas K92 compute $d\tau/dz
(dr/r_{\star})$. Both calculations then proceed to normalise with
respect to $\tau$; this gives identical results \textit{only} when a
single population of galaxies is considered. The value of $r_{\star}$
depends on $\sigma_{\star}$, which in turn varies as a function of the
morphological type considered. We include both ellipticals and S0
galaxies in our analysis, whereas K92 considered only ellipticals.

To obtain constraints on galaxy evolution, we consider the following
scaling relations where

\begin{equation}
n_{\star}(z)=n_{\star}10^{Pz},
\end{equation}
\begin{equation}
L_{\star}(z)=L_{\star}10^{Qz},
\end{equation}
\begin{equation}
\sigma_{\star}(z)=\sigma_{\star}10^{Uz},
\end{equation}

\noindent $n_{\star}$, $L_{\star}$ and $\sigma_{\star}$ are the
characteristic values at zero redshift, and $P$, $Q$ and $U$ are
constants (as in equations (9), (10) and (11) in ORM). Incorporating
these, the IVDF lens-redshift test equation (7) becomes\\[-8mm]

\begin{eqnarray}
\frac{d\tau}{dz dr} & = &
\tau_{\star}\frac{\gamma}{2}\left(\frac{r}{r_{\star}}
\frac{1}{{f_E}^2} \frac{D_S}{D_{LS}}\right)^{\frac{\gamma}{2}(\alpha+1)} \nonumber \\[-2mm]
& & \\[-2mm]
& \times &
\hbox{exp}\left[-\left(\frac{r}{r_{\star}} \frac{1}{{f_E}^2}
  \frac{D_S}{D_{LS}}
\right)^{\frac{\gamma}{2}}{10}^{-Uz\gamma}\right] \nonumber \\[-2mm]
& & \nonumber \\[1mm]
& \times &
\frac{{D_L}^2(1+z)^2}{{r_H}^2E(z)}\frac{r}{{r_{\star}}^2}
10^{[-U\gamma(\alpha+1)+P]z}. 
\nonumber
\end{eqnarray}

Note that Q, the evolution parameter of the luminosity in equation
(9), does not appear in the equation above.

\subsection{Methodology using the measured velocity dispersion function}

In this section we rewrite the differential optical depth as a
function of the measured velocity dispersion function for early-type
galaxies from observations circumventing the use of the Schechter
luminosity function and the Faber-Jackson relation. We now use the
functional form (fitted to SDSS observations) of the MVDF taken from
Sheth et al. (2003) and rewrite it in a form that makes for easy
comparison with the IVDF (equation (4) above):

\begin{equation}
\frac{dn}{d\sigma}=\frac{{n_{\star}}^{\prime}}{\sigma_{\star}^{\prime}}\left(\frac{\sigma}{\sigma_{\star}^{\prime}}\right)^{\alpha^{\prime}-1} \hbox{exp}\left[-\left(\frac{\sigma}{\sigma_{\star}^{\prime}}\right)^{\beta^{\prime}}\right] \frac{\beta^{\prime}}{\Gamma(\alpha^{\prime}/\beta^{\prime})}.
\end{equation}

Substituting and simplifying the expression for the differential
optical depth as done previously, we obtain the MVDF lens-redshift
test equation,\\[-8mm]

\begin{eqnarray}
\frac{d\tau}{dz dr} & = &
{\tau_{\star}}^{\prime}\frac{\beta^{\prime}}{2\Gamma(\alpha^{\prime}/\beta^{\prime})}\left(\frac{r}{r_{\star}^{\prime}} \frac{1}{{f_E}^2} \frac{D_S}{D_{LS}}\right)^{\frac{{\alpha}^{\prime}}{2}} \nonumber \\[-2mm]
& & \\[-2mm]
& \times &
\hbox{exp}\left[-\left(\frac{r}{r_{\star}^{\prime}} \frac{1}{{f_E}^2} \frac{D_S}{D_{LS}}\right)^{\frac{{\beta}^{\prime}}{2}}\right]\frac{{D_L}^2(1+z)^2}{{r_H}^2E(z)}\frac{r}{{r_{\star}^{\prime}}^2}, \nonumber
\end{eqnarray}

\noindent where ${\tau_{\star}}^{\prime} \equiv
16{\pi}^3{n_{\star}}^{\prime}{r_H}^3(\sigma_{\star}^{\prime}/c)^4$ and
$r_{\star} \equiv 4\pi(\sigma_{\star}^{\prime}/c)^2$ are constants.

Once again, we consider the case where the parameters
$n_{\star}^{\prime}$, $L_{\star}^{\prime}$ and
$\sigma_{\star}^{\prime}$ for the MVDF evolve with redshift, following
equations similar to (8), (9) and (10), and on substituting we
find\\[-8mm]

\begin{eqnarray}
\frac{d\tau}{dz dr} & = &
{\tau_{\star}}^{\prime}\frac{\beta^{\prime}}{2\Gamma(\alpha^{\prime}/\beta^{\prime})}\left(\frac{r}{r_{\star}^{\prime}} \frac{1}{{f_E}^2} \frac{D_S}{D_{LS}}\right)^{\frac{{\alpha}^{\prime}}{2}} \nonumber \\[-2mm]
& & \\[-2mm]
& \times &
\hbox{exp}\left[-\left(\frac{r}{r_{\star}^{\prime}} \frac{1}{{f_E}^2} \frac{D_S}{D_{LS}}\right)^{\frac{{\beta}^{\prime}}{2}}10^{-{\beta}^{\prime} U^{\prime}z}\right]\nonumber \\[-2mm]
& & \nonumber \\[1mm]
& \times &
\frac{{D_L}^2(1+z)^2}{{r_H}^2E(z)}\frac{r}{{r_{\star}^{\prime}}^2} 10^{[\alpha^{\prime}- U^{\prime}+P^{\prime}]z}. \nonumber
\end{eqnarray}

There is no reason to assume that $\sigma_{\star}$ and
$\sigma_{\star}^{\prime}$ (or $n_{\star}$ and $n_{\star^{\prime}}$)
evolve differently, therefore, we set $U=U^{\prime}$ (or
$P=P^{\prime}$). We note here that $\sigma_*$ is just a parameter in
the IVDF and the MVDF fits and due to the different functional forms
for the parametrizations of the velocity dispersion function, their
values for the IVDF and MVDF can be and are in fact, found to be quite
different.

\section{Defining the new lens galaxy sample}

Following ORM, our first lens sample is primarily drawn from the
CASTLES (Mu{\~ n}oz et al. 1998) data
base\footnote{http://cfa-www.harvard.edu/castles/index.html} which, at
present, contains 82 class 'A' (certain), 10 class 'B' (likely), and 8
class 'C' (dubious) gravitational lenses, making for a total sample
size of 100 systems. We ignore the 13 class 'B' binary quasars from
the CASTLES lists, eight of which have image separations greater than
4\arcsec\ and would be discarded as likely being cluster-assisted
rather than due to field galaxies. In the remaining 5 lenses nearby
group galaxies are implicated in determining the separations,
therefore we discard them as well.

To get a handle on potential biases in the sample we have grouped the
systems by their discovery technique into three categories: targeted
optical discoveries, which were targets selected based on the lensed
source optical emission and includes such surveys as the first HST
snapshot survey (Maoz et al. 1993) and quasars selected from the
Calan-Tololo (Maza et al. 1996), Hamburg-ESO (Wisotzki et al. 2000)
and SDSS quasar surveys (York et al. 2000); targeted radio
discoveries, which were targets selected based on the lensed source's
radio properties and includes the JVAS/CLASS (Myers et al. 2003), PMN
(Winn et al. 2002), and MG (Bennett et al. 1986) lens searches; and
miscellaneous discoveries for systems discovered either
serendipitously (such as the HST parallel field discoveries;
Ratnatunga, Griffiths \& Ostrander 1999) or based on system properties
other than that of the lensed source and typically that of the lensing
galaxy.

Lenses discovered based on the properties of the source ought not to
harbour a bias in terms of the redshift of the lensing galaxy,
although they suffer from magnification bias. However, systems
discovered because of the lens or surrounding environment will
naturally favour low-redshift lenses. All systems in the
`Miscellaneous Discoveries' fall under this category, which includes
systems discovered because of the properties of the lensing galaxy
such as Q2237+030 (Huchra et al. 1985) and CFRS03.1077 (Crampton et
al. 2002), systems discovered based on properties of the lensing
galaxy's environment such as RXJ0921+4529 (Mu{\~ n}oz et al. 1998),
and systems discovered serendipitously from HST pointings such as the
HST Medium Deep Survey lensing candidates (Ratnatunga, Griffiths \&
Ostrander 1999) and HDFS2232509-603243 (Barkana, Blandford \& Hogg
1999), the latter of which are characterised by deflector emission
that is either comparable to or dominates over the background source.

We also exclude systems inappropriate for our lens model of isolated
and elliptical/S0 lensing galaxies. These include systems with
multiple lensing galaxies of comparable luminosities (and therefore
likely comparable halo masses) such as HE0230-2130 (Wisotzki et
al. 1998), B1359+154 (Myers et al. 1999) and B2114+022 (Augusto et
al. 2001). We also exclude cluster-assisted systems such as Q0957+561
(Young et al. 1980).

Although we are ignoring entire surveys, this ought not to introduce
biases in the sample. These various cuts detailed above leave a total
of 42 systems in our {\bf sample A1} detailed in Table A1 with
complete redshift information (source and lens redshifts). We have
estimated the size of the deflector's critical radius for the
remaining systems using a simple Singular Isothermal Sphere model
(SIS) plus external shear using the {\tt gravlens} software of Keeton
(2001). Relative image positions with respect to the lensing galaxy
were obtained from either the CASTLES compilation or from the
reference in column (11) of Table A1 in the Appendix.  For double
systems, we use the reddest flux ratio measured between lensed
components (typically either HST/F160W or the radio flux ratio if the
system is radio-loud) as the fifth constraint required by the model,
which ought to minimize flux ratio contamination from
microlensing-induced variability.  For systems with ring morphology,
the critical radius of the lens was obtained from the corresponding
model reported in the cited reference.

\begin{figure}
\begin{center}
\includegraphics[width=8cm]{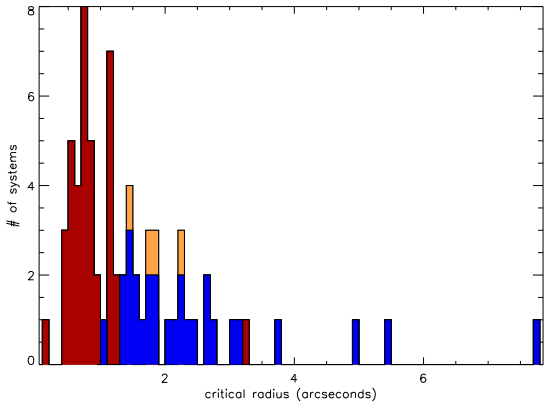}
\caption{Histogram of the critical radii of lenses for the various
samples considered in this work: sample A1 (brown and orange); Bolton
et al. (2006) SLACS sample of 28 lenses (19 confirmed lenses and 9
candidates) (blue). Note that the SLACS lenses have systematically
larger critical radii as a consequence of their spectroscopic
selection from the SDSS.}
\end{center}
\end{figure}

Finally, we add the SLACS lenses (Bolton et al. 2006) to construct our
largest sample comprising seventy systems. The SLACS survey has proven
to be a very efficient HST Snapshot imaging survey for new
galaxy-scale strong lenses. The targeted lens candidates were selected
by Bolton et al. (2006) from the SDSS database of galaxy spectra for
having multiple nebular emission lines at a redshift significantly
higher than that of the target SDSS galaxy. The survey is optimized to
detect bright early-type lens galaxies with faint lensed sources. The
key advantage of this selection technique is that it provides a
homogeneously selected sample of bright early-type lens
galaxies. However, given the size of the fibres in SDSS this sample is
biased toward large separations (the consequences of this bias are
discussed later) compared to other surveys. This is clearly seen in
the histogram plotted in Fig.~1. All the 28 SLACS lenses found to date
are tabulated in Table A2 of the Appendix. Of these lenses 19 are
confirmed with multiple images and the remaining 9 lenses are
candidates. We note here that both the confirmed and unconfirmed
candidates from the SLACS are included in our calculations. Although
the SLACS is a biased sample, we include all 28 lenses in our analysis
to illustrate the effect of better statistics given the success of the
adopted selection strategy. The SLACS selection favours gravitational
lenses that have typically larger Einstein radii (as seen clearly in
the histogram of image separations in Fig.~1) by virtue of the
selection of spectroscopic candidates for imaging follow-up. For the
purposes of the current analysis it clearly strengthens the results of
this work, i.e. biased sampling of the image separation distribution
provides biased values for galaxy evolution parameters. Once the
survey has finished, the selection function will be very well
determined, which will enable more careful use of this sample for
galaxy evolution studies.  This strategy has been extremely successful
so we feel compelled to showcase this sample. SLACS E/S0 lenses appear
to be a random sub-sample of the luminous red galaxies sample of the
SDSS, only skewed toward the brighter and higher surface brightness
systems. While the environments of some of the lenses are complicated
by the existence of nearby galaxies, by and large it is a `clean'
sample where the image separations are determined primarily by a
single elliptical/S0 lens. Including the 28 SLACS lenses to the sample
A1 gives us a final tally of 70 lenses, all with complete information,
that defines our {\bf sample A2}. In Fig.~1, we plot the distribution
of critical radii for observed lenses in sample A1 and the SLACS
lenses, which together constitute our sample A2. Since several samples
will be used in the paper, we list them here for clarity: {\bf Sample
A1 - our updated version of the ORM sample I, with a total of 42
systems; Sample A2 - our new, enlarged sample that contains Sample A1
and 28 new SLACS lenses; Sample B - our mock sample of a 100 lenses
with complete information; and Sample C: our truncated sample A1 with
10\% of the largest separation lenses removed.}

\section{Analysis and results}

In this section, we compare the discriminating power of the MVDF
versus the IVDF, in constraining both cosmological and galaxy
evolution parameters. We study the recovery bias in the extraction of
(i) cosmological constraints with $U = P = 0$ for the various compiled
lens samples, as well as of (ii) galaxy evolution parameters, by
fixing the cosmological parameters. Finally, we assess the impact of
incompleteness of lens data in the recovery of the cosmological
constant.

As noted in section 2, we normalise the lens-redshift probability
distribution with respect to the optical depth $\tau$. Therefore,
parameters which appear simply as multiplying constants in the
distributions do not impact our comparison. Such parameters include
the Hubble radius $r_H=c/H_0$ and the average number density of
lensing galaxies $n_{\star}$ (IVDF) and ${n_{\star}}^{\prime}$
(MVDF). Our comparison of the IVDF and MVDF is not affected by the
value of $f_E$. This parameter relates the velocity dispersion of the
dark matter to that of the stars. TOG set it to $(3/2)^{1/2}$, other
studies (e.g. Narayan \& Bartelmann 1999) suggested using values
smaller than 1. Recent results from the SLACS survey suggest that $f_E
\sim 1$, i.e. the lens model velocity dispersions are fairly close to
the measured stellar velocity dispersion within an effective radius
(Treu et al. 2006). Therefore, we take $f_E = 1$ in this work.  The
default cosmological model is taken to be a Friedmann-Robertson-Walker
($\Omega_{\Lambda} + \Omega_M + \Omega_K = 1$) $\Lambda$-CDM flat
Universe, with $\Omega_M=0.3$, $\Omega_K=0.0$ and
$\Omega_{\Lambda}=0.7$.

The parameters that affect our analysis are the ones defining the
VDFs: $\sigma_{\star}$, $\gamma$, $\alpha$, $\sigma_{\star}^{\prime}$,
$\alpha^{\prime}$, $\beta^{\prime}$. For the IVDF: $\alpha$ is the
faint-end slope in the Schechter luminosity function, and $\gamma$ is
the Faber-Jackson power-law index. For the MVDF: $\alpha^{\prime}$ is
the low-velocity power-law index, and $\beta^{\prime}$ is the
high-velocity exponential cutoff index of the distribution. Following
ORM, the values\footnote{$\sigma_{\star}=225 \hbox{ km s}^{-1}$ is the
characteristic velocity dispersion for elliptical galaxies only; S0
galaxies have a characteristic velocity dispersion of 206 $\hbox{ km
s}^{-1}$. When varying $\sigma_{\star}$ in our calculations, we also
vary the characteristic velocity dispersion of S0 galaxies as
$\sigma_{\star}\times$(206/225).}  $(n_{\star \rm{E}},n_{\star
\rm{S0}},\sigma_{\star},\alpha,\gamma)$ = ($0.0039 h^3
\hbox{Mpc}^{-3}$, $0.0061 h^3 \hbox{Mpc}^{-3}$, $225\hbox{ km
s}^{-1}$, -0.54, 4) are used for the IVDF. Following MKFS fit to the
MVDF, we take\footnote{We note that Sheth et al. (2003) choose a
higher value for ${n_{\star}}^{\prime}$; moreover, $n_{\star
\rm{E}}+n_{\star \rm{S0}}\neq{n_{\star}}^{\prime}$, but the choice of
these parameters does not affect our analysis.}
$({n_{\star}}^{\prime},\sigma_{\star}^{\prime},\alpha^{\prime},\beta^{\prime})$
= ($0.0041 h^3 \hbox{Mpc}^{-3}$, $88.8\hbox{ km s}^{-1}$, 6.5, 1.93).

Note that late-type galaxies can in principle also be incorporated
into the IVDF analysis easily, by simply replacing the Faber-Jackson
relation with the Tully-Fisher relation. We can substitute the
Faber-Jackson exponent $\gamma$ and characteristic $\sigma_{\star}$
with the corresponding Tully-Fisher relation parameters. Although
late-type galaxies are more numerous than early-type galaxies, they
tend to have lower masses and therefore do not contribute
significantly to the total optical depth to lensing. Due to the strong
dependence of the lensing cross section on the velocity dispersion,
this causes late-type galaxies in general to be inefficient
lenses. Besides, late-type galaxies are not included in the
determination of the MVDF from SDSS data. So in this work, for
consistent comparisons we restrict ourselves to elliptical and S0
lenses.

We compute the differential optical depth $(d\tau/dz)/\tau$ for each
individual lens with measured separation and known source redshift for
our two samples: sample A1 (42 lenses) and sample A2 (70 lenses). We
then determine the probability distribution of the redshift of the
lens using equation (7) (IVDF) and equation (13) (MVDF), given the
observed image separation; the measured source redshift; a given
cosmology and galaxy evolution model (in this case assuming no
evolution: $U = P = Q = 0$, $h=0.7$, $\Omega_M=0.3$, $\Omega_K=0.0$
and $\Omega_{\Lambda}=0.7$). If the choice of underlying cosmological
parameters, primarily $\Omega_{\Lambda}$ in this case, corresponds to
the true value, the peak of the probability distribution $z_p$ ought
to be close to the measured lens redshift $z_l$, in a good number of
cases. These lens redshift probability distributions are shown in
Fig.~A2 (in the Appendix) for all the lenses in our sample A2. The
plot illustrates that the MVDF and IVDF yield near identical
probability distributions for the lens redshifts.  Choosing different
values of $\Omega_{\Lambda}$ shifts these inferred probability
distributions: this is illustrated in Fig.~2 for one lens (B0218+357),
for values of $\Omega_{\Lambda} = 0.2, 0.4, 0.7 \hbox{ and } 1.0$
(keeping $\Omega_K=0.0$). However, we notice a systematic effect:
$z_p$ the peak redshift of the probability distribution is skewed
slightly lower for the MVDF compared to the IVDF for almost the entire
sample A2. This can be qualitatively explained as arising due to the
different asymptotic behaviours of the MVDF and the IVDF (see Fig.~A1
in the Appendix).

\begin{figure}
\begin{center}
\includegraphics[angle=270,width=8cm]{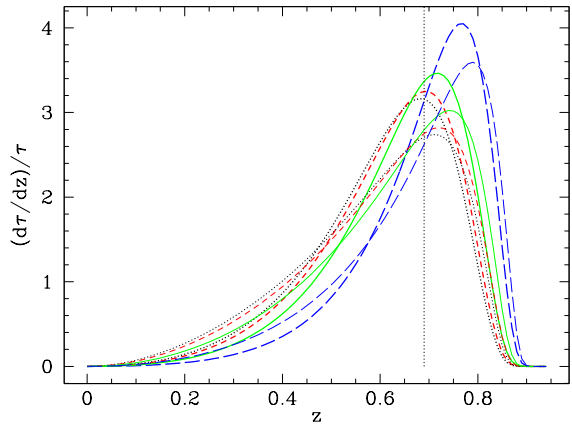}
\caption{Dependence of the lens-redshift distribution on cosmology. We
  show the dependence on the value of $\Omega_{\Lambda}$ explicitly
  for the lens B0218+357 using the MVDF (thick lines) and IVDF (thin
  lines) respectively, for a range of values: $\Omega_{\Lambda}=0.2$
  (black, dotted), $\Omega_{\Lambda}=0.4$ (red, short-dashed),
  $\Omega_{\Lambda}=0.7$ (green, solid) and $\Omega_{\Lambda}=1.0$
  (blue, long-dashed). The vertical dashed line marks the position of
  the observed lens redshift $z_l = 0.69$. The peak $z_p$ of the
  distribution increases as $\Omega_{\Lambda}$ increases from 0.2 to
  1.0, due to the increase in the cosmological comoving volume.}
\end{center}
\end{figure}

\subsection{The maximum likelihood method}

Now, we use the samples to statistically play the game in both
directions: (i) constrain the geometry of the Universe with galaxy
evolution parameters fixed, and (ii) to constrain the galaxy evolution
model with cosmological parameters as knowns.

We use a maximum likelihood estimator in our statistical analysis of
lens redshift distributions. The lens-redshift test equations give the
probability distribution of lens redshifts as $P(z_l | \{X\}, z_s,
r)$, normalised to unity, where $\{X\}=\{\Omega_{\Lambda},U,P\}$ is
the set of cosmological and galaxy evolution model parameters, and
$(z_s, r)$ are source redshift and lens angular critical radius priors
for a given system. The likelihood estimator $L$ for the entire sample
of $N$ systems is then:

\begin{equation}
L(\{X\}) = \prod_{i=1}^N P_i(z_l | \{X\}, z_s, r).
\end{equation}

The quantity $L$ is computed to quantify the consistency of all
measured lens redshifts for the entire ensemble of lens systems for
any given geometry and galaxy evolution model.

We then compute the maximum value of $L$, fixing the galaxy evolution
parameters to obtain constraints on the cosmology
($\{X\}=\{\Omega_{\Lambda}\}$), and then fixing the cosmology to
obtain constraints on galaxy evolution ($\{X\}=\{U,P\}$).

As pointed out by ORM, the lens-redshift test is more sensitive to the
galaxy mass evolution parameter $U$ compared to the galaxy number
evolution parameter $P$. This can be understood by considering the
limit when $P \sim 0$: a negative $U$ decreases the most probable
value for the lens redshift and narrows the probability distribution.
In contrast, the number evolution parameter only affects the peak
value but does not affect the overall shape of the probability
distribution.

\subsubsection{Constraints on cosmology}

We proceed to obtain constraints on $\Omega_{\Lambda}$, keeping galaxy
evolution parameters fixed, using the Friedmann-Robertson-Walker
cosmology, and imposing $\Omega_{\Lambda} + \Omega_M + \Omega_K = 1$
with $\Omega_K=0.0$. Unless otherwise stated, we assume $U=P=Q=0$,
corresponding to the case of no evolution in the galaxy population
either in mass or number.  The lens-redshift test equations (7) and
(13) are used in this instance and the likelihood as described above
is constructed and maximized.

A projection of the likelihood surface along the $\Omega_{\Lambda}$
axis for sample A1 is shown in Fig.~3. In the upper panel the
likelihood function is calculated using the IVDF for sample
A1. Several values of $\sigma_{\star}$ are shown for
completeness. Assuming a value of $\sigma_{\star}=225\hbox{ km
s}^{-1}$ for elliptical galaxies (ORM), we obtain the following limits
on the cosmological constant: $\Omega_{\Lambda}=0.55_{-0.20}^{+0.14}$
at $1\sigma$ confidence. This value is determined by taking the median
as the central value and "ruling out" the leftmost $\sim16$\% and
rightmost $\sim16$\% of the total integral (`median'
method). Alternatively, if we take the mode as the central value and
determine the threshold value of the likelihood for which the integral
under it comprises $\sim68$\% of the total integral (`mode' method),
we obtain a slightly higher value of
$\Omega_{\Lambda}=0.60_{-0.19}^{+0.13}$; this method is similar to
what ORM did, except that they assumed a normal distribution (`normal'
method). Their assumption turns out to be quite reasonable as we
recover the same results ($\Omega_{\Lambda}=0.60_{-0.18}^{+0.12}$)
applying their method to our sample A1. For the IVDF, we note that the
error bars we obtain are much smaller than in ORM: this is certainly
due to the larger number of lens systems in our sample A1 (the ORM
sample I had 15 lenses compared to 42 in our sample A1). Although our
error bars are smaller, we note that the numbers quoted here are for a
single value of $\sigma_{\star} = 225\hbox{ km s}^{-1}$ and these
determinations of $\Omega_{\Lambda}$ using the `mode', `median' and
`normal' methods are entirely consistent with each other within the
errors.

The corresponding results for sample A1 using the MVDF are also shown
in Fig.~3 (lower panel): in this case as well, several values of
$\sigma_{\star}^{\prime}$ are plotted to present the trend
clearly. Assuming $\sigma_{\star}^{\prime}=88.8\hbox{ km s}^{-1}$
(Sheth et al. 2003, MKFS), we obtain a value for the cosmological
constant of $\Omega_{\Lambda}=0.62_{-0.17}^{+0.12}$ using the `median'
method, $\Omega_{\Lambda}=0.67_{-0.16}^{+0.11}$ using the `mode'
method, and $\Omega_{\Lambda}=0.67_{-0.15}^{+0.11}$ using the `normal'
method.  Again, we note that these quoted values are for a single
value of $\sigma_{\star}^{\prime} = 88.8\hbox{ km s}^{-1}$ and once
again the constraints on $\Omega_{\Lambda}$ using these three
different criteria are consistent with each other given the
errors. Even with the improvement of using the MVDF compared to
earlier work, the sensitivity to $\Omega_{\Lambda}$ in the
lens-redshift test is low, as seen clearly by the fact that using the
$\pm 1 \sigma$ range on $\sigma_{\star}^{\prime}$ recovers values of
$\Omega_{\Lambda}$ varying from 0.0 to nearly 1.0 - the full available
range.

\begin{figure}
\begin{center}
\includegraphics[width=8cm]{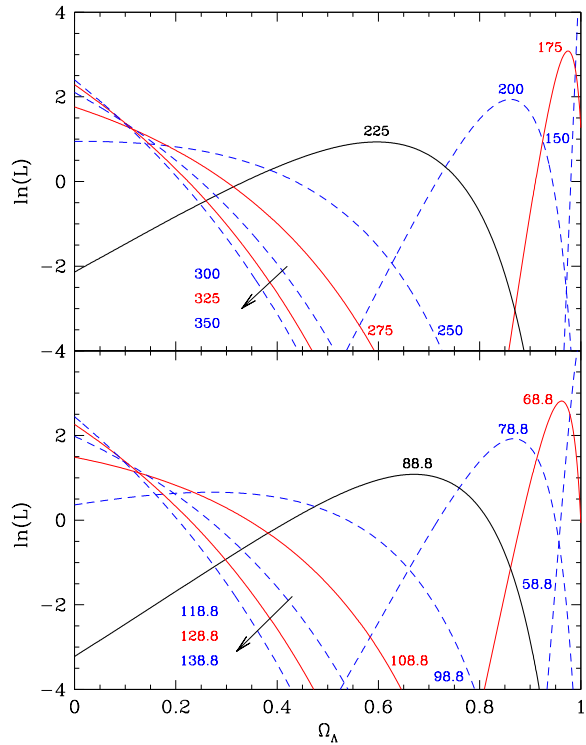}
\caption{The projection of the likelihood surface along the
$\Omega_{\Lambda}$ axis is shown here for several values of the
characteristic velocity dispersion $\sigma_{\star}$ (IVDF; upper
panel) and $\sigma_{\star}^{\prime}$ (MVDF; lower panel) for sample
A1. The numbers denote the characteristic velocity dispersion in
$\hbox{ km s}^{-1}$, increasing in value from the top down in both
panels as indicated by the arrows. For $\sigma_{\star}=225\hbox{ km
s}^{-1}$ (black thick line, upper panel), we obtain
$\Omega_{\Lambda}=0.60_{-0.19}^{+0.13}$. For
$\sigma_{\star}^{\prime}=88.8\hbox{ km s}^{-1}$ (black thick line,
lower panel), we obtain $\Omega_{\Lambda}=0.67_{-0.16}^{+0.11}$. The
likelihood curves are very shallow and consequently the error bars are
rather large. The MVDF and IVDF constraints on $\Omega_{\Lambda}$ are
in good agreement.}
\end{center}
\end{figure}

Unsurprisingly, the recovery of $\Omega_{\Lambda}$ using the MVDF and
the IVDF yields very similar values, as these functions differ only at
the extremely high ($\sigma_{\star}>380\hbox{ km s}^{-1}$) velocity
dispersion tail as shown in Fig.~A1 in the Appendix. A marked
difference between the IVDF and MVDF shows up in the velocity range of
380--400 $\hbox{ km s}^{-1}$, which is characteristic of cD
galaxies. Strong lensing events from such galaxies are difficult to
model, as their position at the centre of clusters causes the events
to be assisted by additional smoothly distributed dark matter in their
vicinity. Since we excluded all such systems in our sample A1, it is
not surprising that the inferred value of the cosmological constant
using the IVDF and MVDF are in good agreement.

We plot the corresponding results for the larger sample A2 in Fig.~4
employing the MVDF. Once again we plot the projection of the
likelihood for various values of $\sigma_{\star}^{\prime}$. For
$\sigma_{\star}^{\prime}=88.8\hbox{ km s}^{-1}$, we now find
$\Omega_{\Lambda}=0.86_{-0.06}^{+0.04}$ for the `median' method;
$\Omega_{\Lambda}=0.87_{-0.06}^{+0.04}$ for the `mode' method and
$\Omega_{\Lambda}=0.87_{-0.06}^{+0.04}$ for the `normal' method.

We find that the recovered value of $\Omega_{\Lambda}$ is higher from
the sample A2 (the mode value is shifted by about 0.25). Sample A2
does include a higher proportion of larger separation lenses (clearly
seen in Fig.~1).  This indicates a potential systematic bias that
skews recovery of $\Omega_{\Lambda}$, that is sensitive to how well
the `true' separation distribution is sampled. To obtain robust
constraints on $\Omega_{\Lambda}$ with the lens-redshift test not only
do we need large samples but we also need lenses that accurately
reflect the true underlying distribution of image separations.  We
note here that the SLACS lenses are included to clearly demonstrate
this bias as their image separations are skewed to larger values as a
consequence of the selection technique.

Four key results emerge from these plots: first, the lens-redshift
test is not very robust in constraining the value of the cosmological
constant with current samples. This was already suggested by K92, but
we demonstrate it more clearly here even with two notable
improvements: a larger sample of lenses and the use of the MVDF. The
likelihood curve is very shallow and consequently the error bars are
rather large. Second, the MVDF and IVDF results are comparable,
therefore the inefficacy of the lens-redshift test does not appear to
stem from systematics arising from the use of the IVDF. Third, is the
notable sensitivity of constraints on $\Omega_{\Lambda}$ to the
parameter $\sigma^{'}_*$.  The value of $\sigma^{'}_*$ emerges in the
fit of a functional form to the observed velocity dispersion function
and depends on the completeness of the measurement, i.e. adequate
sampling of the high and low velocity dispersion tail for observed
galaxies. Finally, inclusion of the SLACS lenses (19 confirmed lenses
+ 9 candidates) with relatively larger separations pushes the
recovered $\Omega_{\Lambda}$ to higher values. The finite number of
lens systems is clearly implicated here as evidenced in the error bars
on $\Omega_{\Lambda}$ and is a key limitation. In conclusion, as we
show in the next section, while a large number of lens systems will go
a ways toward increasing the robustness of this test in the future, it
is crucial to simultaneously sample the separation distribution
uniformly.

\subsubsection{Constraints on galaxy evolution}

We now investigate galaxy evolution using the lens-redshift test, with
$n_{\star}$, $L_{\star}$, $\sigma_{\star}$, $n_{\star}^{\prime}$,
$L_{\star}^{\prime}$ and $\sigma_{\star}^{\prime}$ varying with
redshift according to equations (8), (9) and (10) and their primed
versions. The equations used are (11) and (14). Fixing the
cosmological model to $h=0.7$, $\Omega_M=0.3$, $\Omega_K=0.0$ and
$\Omega_{\Lambda}=0.7$, we determine $U$ and $P$. As outlined in
Section 4.1, the likelihood function is constructed fixing
$\Omega_{\Lambda}$ and then maximizing to obtain constraints on $U$
and $P$.

We obtain constraints on $U$ and $P$ for various samples: the ORM
sample I lenses, sample A1 and sample A2 all of which are plotted in
Fig.~5. We calculate the $U$-$P$ contours for the ORM sample I,
applying our analysis methods and using the MVDF. The MVDF was not
available at the time of the ORM analysis. Our calculation of the $U$
and $P$ parameters using the IVDF is in very good agreement with their
results. The orientation and calibration of the confidence level
contours agree.  Although the contours of all the samples overlap
quite well, the difference in the peak values of $U$ and $P$
determined for our samples A1, A2 and ORM sample I is significant.

The likelihood results in the $U$-$P$ plane for sample A1, using the
IVDF and the MVDF, are shown in the upper and the lower panels of
Fig.~5, respectively. We obtain a maximum in $L$ at \{$U=0.11$,
$P=-1.40$\} for the IVDF and at \{$U=0.10$, $P=-1.24$\} for the MVDF,
again showing no significant dependence on the choice of velocity
dispersion function employed. However, we note that the contours close
along the $U$-axis for the MVDF case compared to the IVDF. Therefore,
using the IVDF to calculate the likelihood lowers the sensitivity to
mass evolution. For our sample A2 the maximum value for $L$ is at
\{$U=0.32$, $P=-1.60$\} for the IVDF and at \{$U=0.28$, $P=-1.57$\}
for the MVDF. For the ORM sample I: the maximum value for $L$ is found
to lie at \{$U=-0.08$, $P=0.44$\} for the IVDF and at \{$U=-0.07$,
$P=0.85$\} for the MVDF. The likelihood peak moves toward increasingly
positive values of U for sample A2 compared to A1 and the ORM sample
I. This indicates a strong sensitivity to the fraction of large
separation lenses. Sample A2 has a larger proportion of those and
therefore predicts stronger mass evolution for the lens ensemble.

The fact that $U$ and $P$ have opposite signs is consistent with mass
conservation: we have either a larger number of lower mass galaxies
($P>0$, $U<0$) or fewer more massive galaxies ($P<0$, $U>0$) in the
past compared to today. However, the case ($P<0$, $U>0$) is in
conflict with the currently accepted hierarchical model of galaxy
formation with bottom-up assembly of structure.

Our primary conclusions on deriving galaxy evolution parameters are:
(i) we reproduce the trends reported by ORM for their sample I when we
use the IVDF as they did; (ii) we find the likelihood peak position to
be insensitive to the choice of IVDF vs MVDF for the ORM sample I;
(iii) we recover $U$ and $P$ values consistent within the errors for
all our samples; (iv) with a larger number of lenses (as in sample A2)
we obtain slightly increased sensitivity to $P$ compared to ORM sample
I; (v) the likelihood peak shifts systematically to higher $U$ values
for sample A2 which contains a higher proportion of large separation
lenses compared to the sample A1. To summarise, there is a notable
observational bias in recovering mass evolution that depends strongly
on how well the underlying true separation distribution is sampled in
detected lenses.

\begin{figure}
\begin{center}
\includegraphics[width=6cm,angle=270]{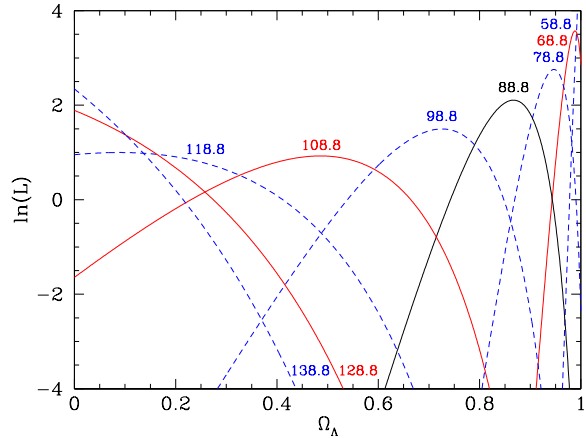}
\caption{The projection of the likelihood surface along the
$\Omega_{\Lambda}$ axis is shown here for several values of the
characteristic velocity dispersion $\sigma_{\star}^{\prime}$ for the
full sample A2 using the MVDF. The numbers denote the characteristic
velocity dispersion in $\hbox{ km s}^{-1}$, increasing in value from
the top down as indicated by the arrows. For For
$\sigma_{\star}^{\prime}=88.8\hbox{ km s}^{-1}$ (black thick line), we
obtain $\Omega_{\Lambda}=0.87_{-0.06}^{+0.04}$. The likelihood curves
are very shallow and consequently the error bars are rather large. The
sample A2 clearly yields values of $\Omega_{\Lambda}$ that are
systematically higher than sample A1.}
\end{center}
\end{figure}

\subsubsection{Investigation of systematic observational biases}

\begin{figure*}
\begin{center}
\includegraphics[width=16cm]{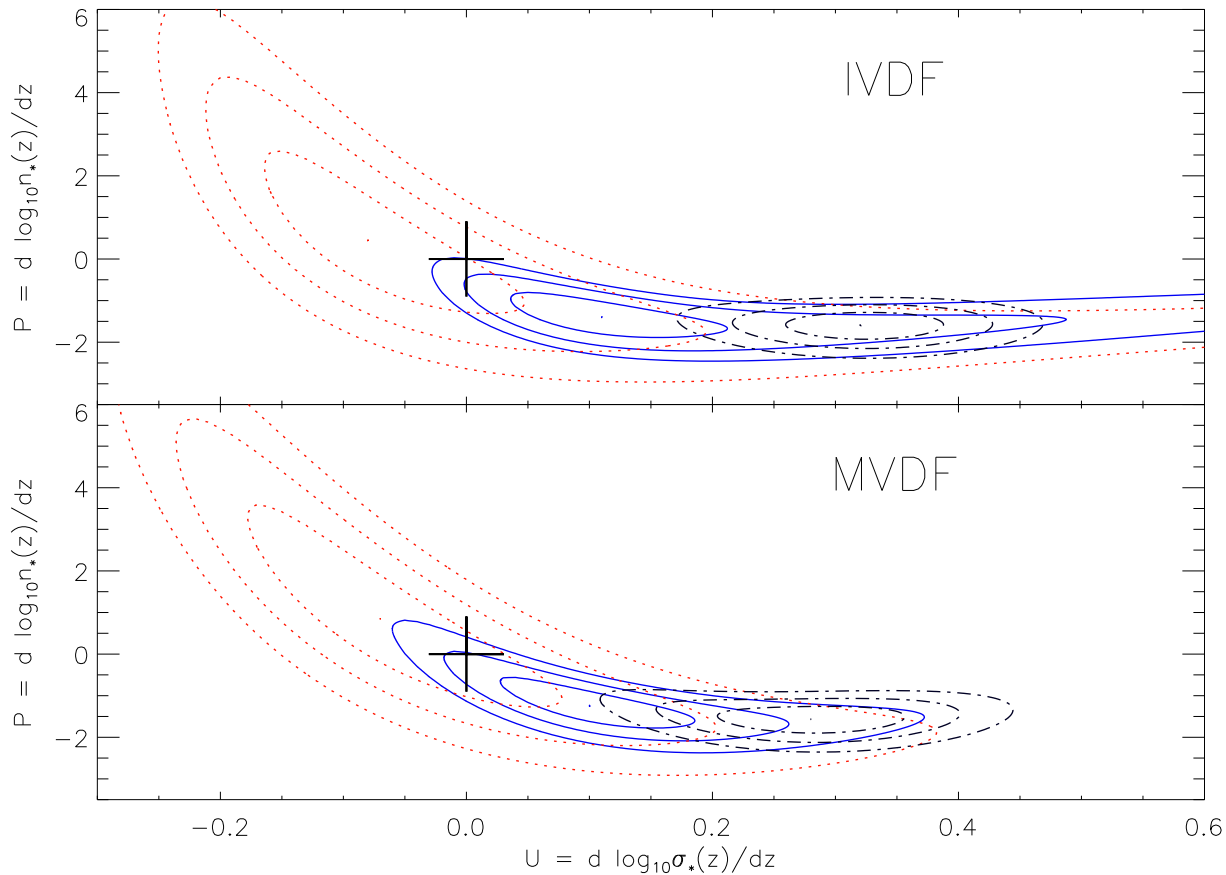}
\caption{Comparison of constraints on galaxy evolution for different
samples: the derived $U$-$P$ contours for sample A1 (blue, solid
lines), sample A2 (black, dot-dashed lines) and the ORM sample I (red,
dashed lines).  In the upper panel, the IVDF was used for the
calculation and in the lower panel the MVDF. The three contours shown
for each sample are the $1\sigma$, $2\sigma$ and $3\sigma$ confidence
levels and the cross marks the no-evolution locus $(U=0,P=0)$. While
there is some degree of overlap for the various samples it is clear
that the peak value of $U$ -- the mass evolution parameter -- shifts
toward more positive values consistent with sample A2 having a larger
proportion of more massive lenses, in good agreement with the fact
that sample A2 does have a higher fraction of large separation
lenses. The error bars on the ORM sample I are larger since it has
only 15 systems whereas sample A1 has 42 systems and sample A2 70
systems. We note that within the errors the values of $U$ and $P$
obtained for different samples are in good agreement.}
\end{center}
\end{figure*}

Unbiased lens surveys are needed to sample uniformly the full
distribution of separations (Kochanek 1993) in order to apply the
lens-redshift test to constrain galaxy evolution parameters as found
above. If the sample is slightly skewed toward larger separations,
biased values of the galaxy evolution parameters are retrieved. To
investigate this issue further, we create a mock sample of a hundred
lenses ({\bf sample B}), all with complete information (i.e. source
redshift, lens redshift, and image separation) known assuming no
evolution, to try and understand the observational biases that likely
affect our analysis. We randomly assign the source redshift from a
normal distribution of redshifts centred at $z=2$ and with a
dispersion of 1. We then randomly assign the angular critical radius,
as half of the lens separation, from the probability distribution of
image separations given by Kochanek (1993), where we set $\gamma=4$
and $\alpha=-1$ in their equation (4.10) for a flat Universe. We
calculate the differential optical depth distribution for each lens,
using equation (7), and randomly pick a lens redshift from it.

\begin{figure*}
\begin{center}
\includegraphics[width=16cm]{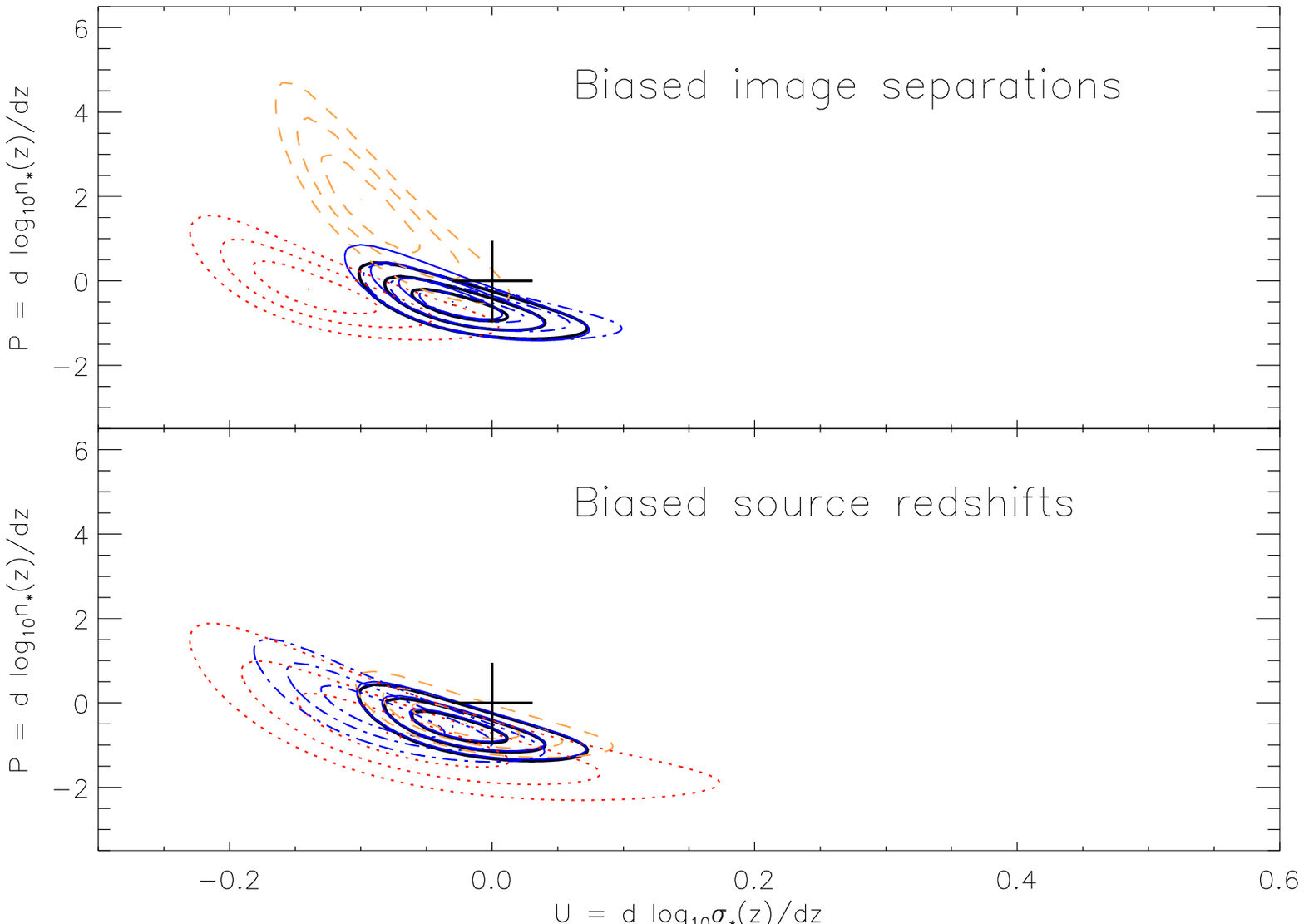}
\caption{Effect of biased lens samples and the recovery of $U$ and $P$
illustrated for the IVDF case. Upper panel [Biased image separations]:
$U$-$P$ contours for our full mock sample B and several subsamples
selected by cuts in image separations (i) lowest $90\%$ separations
(blue, dot-dashed thin lines), (ii) lowest $58\%$ separations (red,
dotted thin lines), (iii) largest $90\%$ separations (blue, solid thin
lines), (iv) largest $58\%$ separations (orange, dashed thin
lines). Lower panel [Biased source redshifts]: $U$-$P$ contours for
subsamples now cut on the basis of source redshifts (i) lowest $90\%$
$z_s$ (blue, dot-dashed thin lines), (ii) lowest $58\%$ $z_s$ (red,
dotted thin lines), (iii) highest $90\%$ $z_s$ (blue, solid thin
lines), (iv) highest $58\%$ $z_s$ (orange, dashed thin lines). The
full mock sample B results (black, solid thick lines) are shown in
both panels for comparison.  The three contours shown for each sample
are the $1\sigma$, $2\sigma$ and $3\sigma$ contour likelihood lines
and the cross marks the location of the no-evolution locus
$(U=0,P=0)$. A clear systematic bias is introduced on selection by
lens separation. In particular, the velocity dispersion evolution
parameter dramatically shifts from near-zero values ($U\sim0$) to
negative values ($U\sim-0.1$), when the largest separation lenses are
removed from the sample.}
\end{center}
\end{figure*}

\begin{figure*}
\begin{center}
\includegraphics[width=16cm]{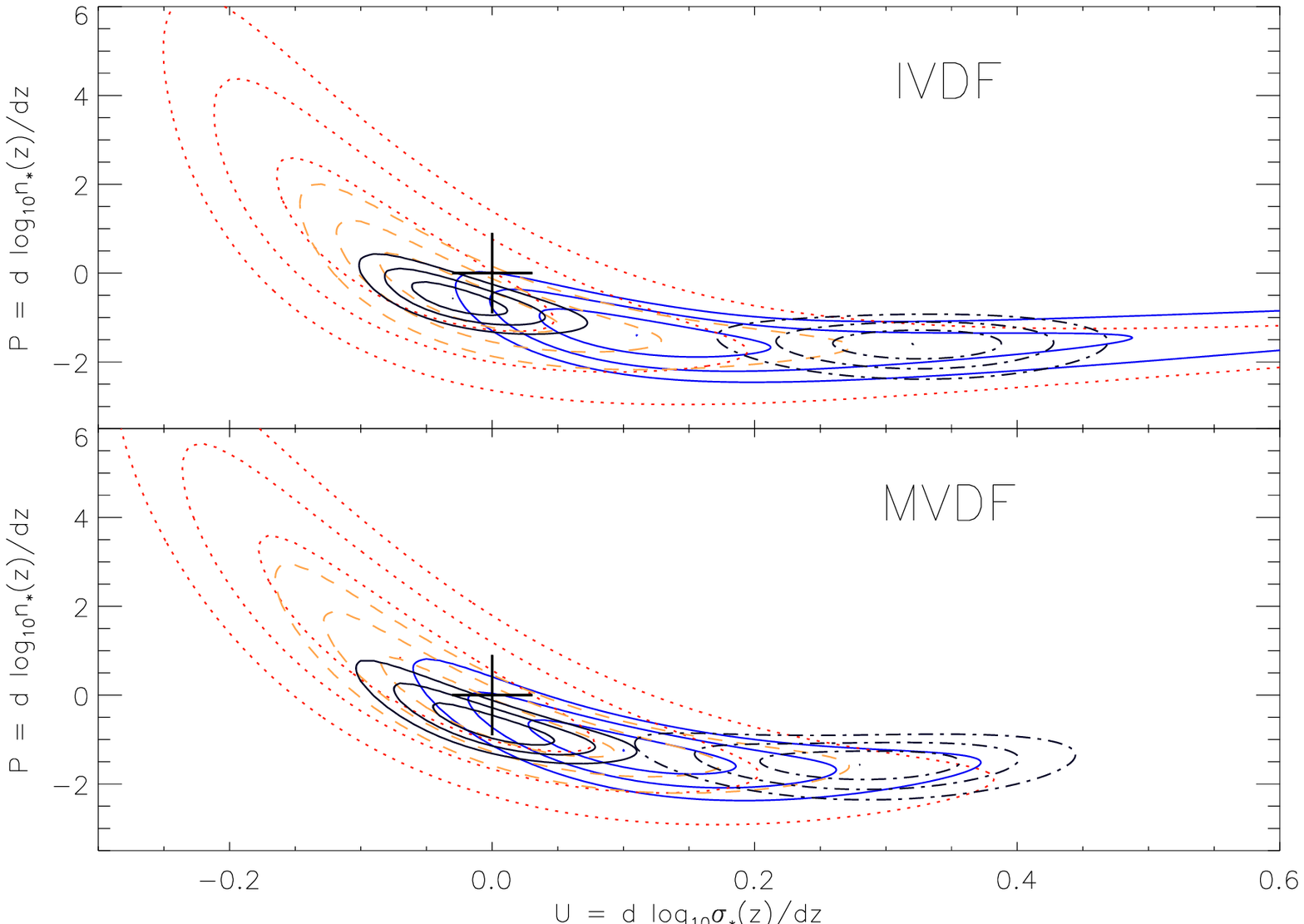}
\caption{Illustrating the systematic bias arising from poorly sampling
the true image separation distribution: the $U$-$P$ contours for ORM
sample I (red, dotted lines) and our samples A1 (blue, solid lines),
A2 (black, dot-dashed lines), C (orange, dashed lines), and B (black,
solid lines) are shown for the IVDF (upper panel) and MVDF (lower
panel). The three contours shown for each sample are the $1\sigma$,
$2\sigma$ and $3\sigma$ confidence levels, with the cross marking
$(U=0,P=0)$ denoting the no-evolution case. The systematic bias
induced in our sample A2 arises from a larger fraction of large image
separation lenses.}
\end{center}
\end{figure*}

For the full mock catalogue (sample B), the maximum value for the
likelihood $L$ is obtained at \{$U=-0.03$, $P=-0.57$\}. The input
$(U=0,P=0)$ parameters values are not exactly recovered due to
finite-sample variance. Several subsamples of sample B were then
evaluated, after cutting the sample based on source redshifts and
image separations. Creating a subsample of 90 lenses discarding the 10
highest source redshift systems, we find that the maximum value of the
likelihood $L$ shifts to \{$U=-0.09$, $P=-0.21$\}. Then for a
subsample of 90 lenses generated by discarding the 10 lowest source
redshift systems we find that the peak is at \{$U=-0.03$, $P=-0.54$\}.

To examine the additional sensitivity to the number of lenses as well,
we now construct a subsample from sample B of 58 lenses with the
lowest source redshifts and find \{$U=-0.07$, $P=-0.80$\} and for the
subsample of 58 lenses with the highest source redshifts the evolution
parameters are found to be \{$U=-0.02$, $P=-0.48$\}.  These results
are shown in the lower panel of Fig.~6: no significant systematic bias
is introduced on selection by source redshift. The only difference
seen in these subsamples is the effect of the variation in the number
of systems: the contours are more extended for the 2 cases with 58
systems compared to the cases with 90 systems.

We then cull sample B on the basis of image separations and the
results for these subsamples are shown in the upper panel of Fig.~6.
The key systematic that we study in further detail is the role of
biased sampling of the image separation distribution. The mock
catalogue generated above was now cut based on lens angular critical
radius $r$. Once again biased subsamples were generated to
preferentially sample larger and smaller separation systems. First, we
constructed a subsample of 90 systems discarding 10 smallest
separation systems: for this instance the maximum value of the
likelihood lies at \{$U=-0.03$, $P=-0.49$\}. Picking now a further 90
systems discarding the 10 largest separation systems, we find a
different maximum at \{$U=-0.02$, $P=-0.62$\}.  Similarly, making a
more extreme selection, we pick 58 systems from the mock discarding 42
of the largest separation systems. This is our extreme biased sample
skewed to small separations. For this subsample we find \{$U=-0.14$,
$P=-0.19$\}. Finally, for a mock with 58 of the largest separation
lenses (this constitutes our extreme biased sample toward large
separations) we find \{$U=-0.10$, $P=1.92$\}. Our analysis clearly
indicates the presence of a systematic bias introduced on selection by
lens separation. This effect is especially seen clearly in the smaller
subsamples (with 58 systems). In particular, the velocity dispersion
evolution parameter dramatically shifts from near-zero values
($U\sim0$) to negative values ($U\sim-0.1$), when the highest
separation lenses are removed from the sample.  We see clearly from
Fig.~6 that lens data comprising biased sampling of the underlying
image separation distribution introduce a systematic shift in the
recovered values of the galaxy evolution parameters, whereas data with
biased source redshifts yield unbiased estimates of $U$ and $P$.

Galaxy evolution parameters are thus extremely sensitive to
observational biases in the separation distribution of lens
systems. For ground based optical surveys and high resolution HST
surveys there are optimal separations that are detected. Lens systems
found in ground based surveys are likely skewed to larger separations
than those found in HST surveys.

Having narrowed the plausible source of the systematic bias, we re-do
the analysis making a similar cut on our observed lens sample A1 to
verify our finding. We remove the five (10\%) largest image separation
lenses thus creating the {\bf sample C}. We now compare the recovery
of $U$ and $P$ for this biased sample with samples A1 and A2, sample B
and the ORM sample I. The maximum value of $L$ is obtained for sample
C at \{$U=-0.03$, $P=-0.51$\} (IVDF) and at \{$U=-0.01$, $P=-0.40$\}
(MVDF). In Table 1 we list the positions of the peak values of the
likelihood function in the $U$-$P$ plane, for our samples A1, A2, B,
ORM sample I and sample C, for the IVDF and MVDF. The full results are
shown in Fig.~7. The resultant trend clearly demonstrates the strong
bias now replicated with cuts in the data introduced by artificially
removing large image separation systems. Our result that
incompleteness in image separations is a serious current limitation in
using strong lensing statistics has also been pointed out by Oguri
(2005) and Oguri, Keeton \& Dalal (2005).

\begin{table}
\centering
\begin{tabular}{lllll}
\hline
Sample       & U=$\hbox{d log}_{10} \sigma_{\star}(z)/dz$ & P=$\hbox{d log}_{10} n_{\star} (z)/dz$\\
\hline
A2 (IVDF)    & +0.32 & -1.60\\
A1 (IVDF)    & +0.11 & -1.40\\
B (IVDF)     & -0.03 & -0.57\\
C (IVDF)     & -0.03 & -0.51\\
ORM I (IVDF) & -0.08 & +0.44\\
\hline
A2 (MVDF)    & +0.28 & -1.57\\
A1 (MVDF)    & +0.10 & -1.24\\
B (MVDF)     & +0.00 & -0.71\\
C (MVDF)     & -0.01 & -0.40\\
ORM I (MVDF) & -0.07 & +0.85\\
\hline
\end{tabular}
\caption{Positions of the peak values of the likelihood function in
the U-P plane, for samples A2, A1, B, C, and ORM I, IVDF and MVDF, in
order of U-peak position. When comparing samples A2, A1 and C, we
confirm the same trend we observed with the mock subsamples: the
velocity dispersion evolution parameter shifts to less positive values
as we remove the highest separation lenses.}
\end{table}

\section{The effect of incomplete lens data on the retrieval of cosmological
parameters}

Previous lens-redshift test analyses have differed on how to handle
systems with incomplete redshift information. K96 included an estimate
of the probability of failing to measure a system's lens redshift for
systems lacking such a measurement. ORM take a more pragmatic approach
by discarding all systems with $z_s > 2.1$, arguing that systems below
that redshift are mostly complete. The former approach is made
difficult by the many variables that can prevent a successful redshift
measurement (surface brightness of the lensing galaxy, galaxy contrast
with respect to the magnified source images, observing conditions
during an actual measurement attempt), while the latter approach
ignores higher-redshift systems that do have complete redshift
information. Such systems are likely to show the strongest sensitivity
to cosmological or evolution effects that are sought after in the
first place.

The approach we adopt here is to marginalise over systems with
incomplete redshift information using nested Monte Carlo
simulations. Let $N_c$ be the number of systems with complete redshift
information and $N_u$ be the number of systems with unmeasured lensing
redshifts. For a given parameter set $\{X\}$, we can assign lens
redshifts for the $N_u$ sample by drawing from $P(z_l | \{X\}, z_s,
r)$ which gives a sample of lens redshifts $\{z_{l,u}\}$. With
$\{z_{l,u}\}$ fixed, we obtain the absolute likelihood $L(\{X\})$ for
the combined $N_c + N_u$ sample. The procedure is then repeated
$N_{MC}$ times with each iteration using a different set of
$\{z_{l,u}\}$. This gives an average absolute likelihood $<L(\{X\})>$
and a corresponding scatter $\delta L(\{X\})$ for the given set of
model parameters $\{X\}$. The scatter in the absolute likelihood
estimate shrinks to zero as $N_u \rightarrow 0$, and can be
interpreted as a measure of the uncertainty in the absolute likelihood
because of the incomplete sample. The entire procedure can then be
repeated for a different set of model parameters.

We argue that this is an attractive method for several reasons. First,
it does not ignore existing redshift information for any system,
either within the complete or the incomplete sample. This results in
as large a sample size as possible and helps to minimize small-number
effects that are traditionally present in lensing statistics. Second,
the question of handling biases present in the incomplete sample is
made objectively by marginalising over the entire sample rather than
imposing an artificial cut on, say, the source redshifts. And third,
it allows one to quantify the effects that the incomplete sample has
on the accuracy of likelihood analysis through $\delta L(\{X\})$. This
last point can be used to explore how the precision of the model
parameters can be measured by future changes in either the complete or
incomplete sample size.

We performed a nested Monte Carlo simulation of our sample A1
($N_c=42$), adding a set of ten mock lens systems with known image
separations, source redshifts and unknown lens redshift ($N_u=10$), to
make up a total of 52 lens systems. We fixed all galaxy evolution
parameters ($U=P=Q=0$, $\sigma_{\star}=225\hbox{ km s}^{-1}$) and
varied cosmological parameters, after setting $\Omega_{\Lambda} +
\Omega_M + \Omega_K = 1$ with $\Omega_K=0.0$. Therefore, the parameter
set was taken to be $\{X\}=\{\Omega_{\Lambda}\}$. A projection of the
(un-normalised) likelihood surface along the $\Omega_{\Lambda}$ axis
is shown in Fig.~8 for the IVDF case: we show the results for
$\Omega_{\Lambda}$ values steps of of 0.1 only, and for
$N_{MC}=100$. The blue line is the result for the sample A1. For each
fixed value of $\Omega_{\Lambda}$, each of the points represents the
value of the likelihood for one of the $N_{MC}=100$ different possible
sets of the combined $N_c + N_u$ sample. Approximating the likelihood
distributions with Gaussian functions of mean $\mu=<L(\{X\})>$ and
dispersion $\sigma=\delta L(\{X\})$, the average absolute likelihood
$<L(\{X\})>$ for the combined $N_c + N_u$ sample follows the pattern
given by the complete system (blue line), being almost flat between
$\Omega_{\Lambda}=0.0$ and $\Omega_{\Lambda}=0.7$. Therefore adding
even a small number of systems with incomplete information reduces
further the sensitivity to $\Omega_{\Lambda}$.\footnote{The likelihood
distributions have very large dispersions indicating the lack of
robustness in determined values of the cosmological constant.  Below
we enumerate some typical values of the mean $\mu=<L(\{X\})>$ and
dispersion $\sigma=\delta L(\{X\})$ in the likelihood stepping through
a grid of $\Omega_{\Lambda}$ values that illustrates this point:
$\Omega_{\Lambda}=0.0$: $\mu\simeq-12.6$, $\sigma\simeq7.9$;
$\Omega_{\Lambda}=0.1$: $\mu\simeq-10.0$, $\sigma\simeq4.7$;
$\Omega_{\Lambda}=0.2$: $\mu\simeq-11.2$, $\sigma\simeq7.4$;
$\Omega_{\Lambda}=0.3$: $\mu\simeq-11.8$, $\sigma\simeq8.0$;
$\Omega_{\Lambda}=0.4$: $\mu\simeq-11.5$, $\sigma\simeq8.6$;
$\Omega_{\Lambda}=0.5$: $\mu\simeq-10.2$, $\sigma\simeq7.8$;
$\Omega_{\Lambda}=0.6$: $\mu\simeq-9.7$, $\sigma\simeq8.1$;
$\Omega_{\Lambda}=0.7$: $\mu\simeq-13.3$, $\sigma\simeq9.9$;
$\Omega_{\Lambda}=0.8$: $\mu\simeq-12.2$, $\sigma\simeq5.5$;
$\Omega_{\Lambda}=0.9$: $\mu\simeq-18.9$, $\sigma\simeq5.3$;
$\Omega_{\Lambda}=1.0$: $\mu\simeq-35.3$, $\sigma\simeq6.9$.}

\begin{figure}
\begin{center}
\includegraphics[angle=270,width=8cm]{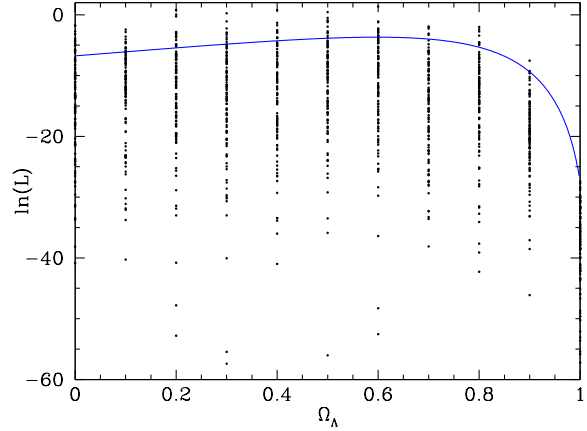}
\caption{The projection of the (un-normalised) likelihood surface
along the $\Omega_{\Lambda}$ axis is shown here for hundred
independent realizations of 52 lens systems (IVDF case), comprising 42
systems from sample A1 and 10 mock systems. All galaxy evolution
parameters being fixed ($U=P=Q=0$, $\sigma_{\star}=225\hbox{ km
s}^{-1}$), each dot represents the value of the likelihood L for a
particular value of $\Omega_{\Lambda}$ and for a particular set of 52
lenses. The blue line is the result for sample A1 (42 lenses)
only. For each value of $\Omega_{\Lambda}$, the distribution of 100
likelihood values can be approximated by a Gaussian function with a
mean $\mu=<L(\{X\})>$ and a dispersion $\sigma=\delta L(\{X\})$. }
\end{center}
\end{figure}

The effect of incomplete information is to further dilute the efficacy
of constraints on cosmological parameters. While we have argued here
that the lens-redshift test with a small lens sample with complete
information is insufficient, we further find even with a small number
of systems in a large sample with incomplete information, we lose
sensitivity to $\Omega_{\Lambda}$.

\section{Conclusions and Discussion}

We investigate the lens-redshift test to assess its robustness in
constraining cosmological and galaxy evolution parameters. We apply
the test to a much improved lens sample compared to earlier work by
K92 and ORM. Moreover, we also use the observationally determined
velocity dispersion function (MVDF), instead of relying on the
IVDF. MKFS also used the MVDF, but they applied it to the ODTL test,
which is more affected by observational biases - mainly the
magnification bias - than the lens-redshift test considered here.
Finally, we develop a new nested Monte Carlo analysis to quantify the
effects of incompleteness on the accuracy of retrieving
$\Omega_{\Lambda}$.

Our results suggest that the lens-redshift test is not particularly
robust in the determination of either cosmological parameters or
galaxy evolution parameters with the currently available samples.  We
conclude this after careful analysis of 70 lens systems and generating
several mock catalogues.  First, we fix galaxy evolution parameters to
constrain $\Omega_{\Lambda}$: in this instance very weak constraints
are obtained. Moreover, despite using the MVDF for the first time in
this test our results do not differ significantly from earlier
work. When we do the converse, i.e. fix the cosmology and look for
constraints on galaxy evolution, we find that the results on the
evolution parameters are too sensitive to the choice of sample,
implying a very strong dependence on the observational bias introduced
by lens separations. Finally, the limit of the precision with which
the value of the cosmological constant can be determined due to lack
of complete information has been assessed. We find that even a small
number of systems with incomplete information in a large sample can
further reduce the significance of the already weak constraints on the
cosmological constant. In fact, systems with incomplete information
add more noise than signal. For the purposes of constraining
cosmological parameters incomplete-redshift information systems are
best excluded.  With the small number of systems available at the
present time, such a strong cut is not feasible, however with the
expected large number of new lenses from future surveys the statistics
will permit stricter selection of optimal systems.

The lens-redshift test is clearly affected by lens selection
effects. An obvious observational strategy for the future would be to
observe hundreds of new lenses, that fairly sample the full
distribution of separations. Such samples are expected from the large
area surveys to be performed by future instruments like SNAP and the
LSST. These large samples with hundreds/thousands of lenses at several
redshifts would allow us to better quantify the lens sample selection
bias. Moreover, ideally lenses for use in the lens-redshift test need
to be relatively ``clean'', that is, they should not belong to groups,
where presence of additional deflectors/nearby galaxies could affect
the image separation and therefore provide skewed lens image
separation distributions that will in turn bias results. It is
becoming increasingly clear from the study of individual lens
environments that there are clearly no truly isolated lenses. However,
what is important from the point of view of the lens-redshift test is
that the neighbouring perturbers are not massive enough to
significantly alter the image separations to within observational
positional accuracies. Obviously these accuracies depend on whether
space based data or ground based data is available for new lens
systems. From large proposed future surveys which all involve deep
imaging, lens systems with many perturbers need to be culled. It
appears from simulations that systems that likely require of the order
of 10 - 20\% external shear are still viable candidates for the
lens-redshift test (Oguri, Keeton \& Dalal 2005) .

Future lens samples should ideally include both early and late type
galaxies and span a large redshift range, in order to constrain galaxy
evolution parameters. Finally, all lens systems to be useful should
have complete information, since even a small fraction of incomplete
systems would significantly decrease the efficacy of constraints on
parameters. In the near future, hundreds/thousands of new lens systems
will be discovered by these upcoming new instruments. Simultaneously
with many on-going and planned ambitious surveys to study galaxy
evolution, progress is likely to come from better knowledge of galaxy
evolution. The lens-redshift test, which is currently unable to give
decent constraints on cosmology and galaxy evolution because of poor
statistics, could eventually prove to be a very profitable means to
constrain cosmological parameters and galaxy evolution models
robustly.

\section*{Acknowledgments}

Nick Morgan is thanked for many useful and insightful comments and for
suggesting the examination of incomplete information on the retrieval
of cosmological constraints. The authors also wish to thank Eran Ofek
and Paolo Coppi for helpful comments during the course of this work
and Chuck Keeton for thoughtful comments on the manuscript. PN
acknowledges discussions with Chris Kochanek on the issue of dusty
lenses.

\vspace{18cm}
\appendix
\section[]{Tables and additional figures}

\label{LensTableApp}

\begin{table*}
\centering
\begin{tabular}{lllllllllll}
 \hline
Number & Name          & R.A.         &  Dec.          & $z_{s}$ & $z_{l}$ & $r$     & Grade & $N_{im}$ & Samples & References \\
(1)    & (2)           & (3)          & (4)            & (5)     & (6)     & (7)     & (8)   & (9)      & (10)    & (11)\\
\hline
1      & HE0047-1756   & 00:50:27.83  & $-$17:40:8.8   & $1.67$  & $0.41$  & $0.773$ & A     & 2        & A1,C,O   & 1,2\\
2      & Q0142-100     & 01:45:16.5   & $-$09:45:17    & $2.72$  & $0.49$  & $1.205$ & A     & 2        & A1,C,O   & 3,45\\
3      & QJ0158-4325   & 01:58:41.44  & $-$43:25:04.20 & $1.29$  & $0.28$  & $0.615$ & A     & 2        & A1,C,O   & 3,4\\
4      & HE0435-1223   & 04:38:14.9   & $-$12:17:14.4  & $1.69$  & $0.46$  & $1.208$ & A     & 4        & A1,C,O   & 2,5,47\\
5      & HS0818+1227   & 08:21:39.1   & $+$12:17:29    & $3.12$  & $0.39$  & $1.199$ & A     & 2        & A1,C,O   & 6\\
6      & SDSS0903+5028 & 09:03:34.92  & $+$50:28:19.2  & $3.61$  & $0.39$  & $1.451$ & A     & 2        & A1,O     & 7\\
7      & RXJ0911+0551  & 09:11:27:50  & $+$05:50:52.0  & $2.80$  & $0.77$  & $1.104$ & A     & 4        & A1,C,O   & 8,9\\
8      & SBS0909+523   & 09:13:01.05  & $+$52:59:28.83 & $1.38$  & $0.83$  & $0.534$ & A     & 2        & A1,C,I,O & 10\\
9      & SDSS0924+0219 & 09:24:55.87  & $+$02:19:24.9  & $1.52$  & $0.39$  & $0.866$ & A     & 4        & A1,C,O   & 2,11\\
10     & FBQ0951+2635  & 09:51:22.57  & $+$26:35:14.1  & $1.24$  & $0.26$  & $0.541$ & A     & 2        & A1,C,I,O & 3,12,45\\
11     & BRI0952-0115  & 09:55:00.01  & $-$01:30:05.0  & $4.50$  & $0.63$  & $0.525$ & A     & 2        & A1,C,O   & 2,3,13,45\\
12     & J1004+1229    & 10:04:24.9   & $+$12:29:22.3  & $2.65$  & $0.95$  & $0.770$ & A     & 2        & A1,C,O   & 14\\
13     & LBQS1009-0252 & 10:12:15.71  & $-$03:07:02.0  & $2.74$  & $0.87$  & $0.779$ & A     & 2        & A1,C,O   & 2,15,46\\
14     & Q1017-207     & 10:17:24.13  & $-$20:47:00.4  & $2.55$  & $0.86$  & $0.461$ & A     & 2        & A1,C,O   & 2,3,16\\
15     & FSC10214+4724 & 10:24:34.6   & $+$47:09:11    & $2.29$  & $0.96$  & $0.820$ & A     & 2E       & A1,C,O   & 3,17\\
16     & HE1104-1805   & 11:06:33.45  & $-$18:21:24.2  & $2.32$  & $0.73$  & $1.724$ & A     & 2        & A1,O     & 18\\
17     & PG1115+080    & 11:18:17.00  & $+$07:45:57.7  & $1.72$  & $0.31$  & $1.144$ & A     & 4        & A1,C,I,O & 19,20\\
18     & RXJ1131-1231  & 11:31:51.6   & $-$12:31:57    & $0.66$  & $0.30$  & $1.817$ & A     & 4        & A1,O     & 21\\
19     & SDSS1138+0314 & 11:38:03.70  & $+$03:14:58.0  & $2.44$  & $0.45$  & $0.664$ & A     & 4        & A1,C,O   & 22,23\\
20     & SDSS1155+6346 & 11:55:17:35  & $+$63:46:22.0  & $2.89$  & $0.18$  & $0.897$ & A     & 2        & A1,C,O   & 22\\
21     & SDSS1226-0006 & 12:26:08.10  & $-$00:06:02.0  & $1.12$  & $0.52$  & $0.634$ & A     & 2        & A1,C,O   & 23\\
22     & SDSS1335+0118 & 13:35:34.79  & $+$01:18:05.5  & $1.57$  & $0.44$  & $0.780$ & A     & 2        & A1,C,O   & 23\\
23     & Q1355-2257    & 13:55:43.38  & $-$22:57:22.9  & $1.37$  & $0.70$  & $0.646$ & A     & 2        & A1,C,O   & 2,45\\
24     & SBS1520+530   & 15:21:44.83  & $+$52:54:48.6  & $1.86$  & $0.72$  & $0.743$ & A     & 2        & A1,C,I,O & 24\\
25     & WFI2033-4723  & 20:33:42.08  & $-$47:23:43.0  & $1.66$  & $0.66$  & $0.980$ & A     & 4        & A1,C,O   & 2\\
26     & HE2149-2745   & 21:52:07.44  & $-$27:31:50.2  & $2.03$  & $0.60$  & $0.857$ & A     & 2        & A1,C,I,O & 25,26,45\\
27     & B0218+357     & 02:21:05.483 & $+$35:56:13.78 & $0.94$  & $0.69$  & $0.169$ & A     & 2ER      & A1,C,I,R & 27,28\\
28     & MG0414+0534   & 04:14:37.73  & $+$05:34:44.3  & $2.64$  & $0.96$  & $1.187$ & A     & 4E       & A1,C,R   & 29,30\\
20     & B0712+472     & 07:16:03.58  & $+$47:08:50.0  & $1.34$  & $0.41$  & $0.716$ & A     & 4        & A1,C,I,R & 31\\
30     & MG0751+2716   & 07:51:41.46  & $+$27:16:31.35 & $3.20$  & $0.35$  & $0.402$ & A     & R        & A1,C,R   & 30\\
31     & B1030+074     & 10:33:34.08  & $+$07:11:25.5  & $1.54$  & $0.60$  & $0.514$ & A     & 2        & A1,C,I,R & 31\\
32     & B1152+200     & 11:55:18.3   & $+$19:39:42.2  & $1.02$  & $0.44$  & $0.807$ & A     & 2        & A1,C,I,R & 32\\
33     & B1422+231     & 14:24:38.09  & $+$22:56:00.6  & $3.62$  & $0.34$  & $0.779$ & A     & 4E       & A1,C,R   & 20,33\\
34     & MG1549+3047   & 15:49:12.37  & $+$30:47:16.6  & $1.17$  & $0.11$  & $1.150$ & A     & R        & A1,C,I,R & 34,35\\
35     & PMN1632-0033  & 16:32:57.68  & $-$00:33:21.1  & $3.42$  & $1.17$  & $0.731$ & B     & 2R       & A1,C,R   & 2,36\\
36     & MG1654+1346   & 16:54:41.83  & $+$13:46:22.0  & $1.74$  & $0.25$  & $0.982$ & A     & R        & A1,C,I,R & 37\\
37     & PKS1830-211   & 18:33:39.94  & $-$21:03:39.7  & $2.51$  & $0.89$  & $0.471$ & A     & 2ER      & A1,C,R   & 38\\
38     & B1933+503     & 19:34:30.95  & $+$50:25:23.6  & $2.62$  & $0.76$  & $0.506$ & A     & 2R       & A1,C,R   & 49\\
39     & Q0047-2808    & 00:49:41.89  & $-$27:52:25.7  & $3.60$  & $0.48$  & $1.163$ & A     & 4ER      & A1,C,M   & 40\\
40     & HST14113+5211 & 14:11:19.60  & $+$52:11:29.0  & $2.81$  & $0.47$  & $2.260$ & A     & 4        & A1,M     & 9,10\\
41     & HST14176+5226 & 14:17:36.61  & $+$52:26:40.0  & $3.40$  & $0.81$  & $3.250$ & A     & 4        & A1,M     & 41\\
42     & HST15433+5352 & 15:43:20.9   & $+$53:51:52    & $2.09$  & $0.50$  & $1.176$ & A     & 2R       & A1,C,M   & 41\\
43     & HE0512-3329   & 05:14:10.78  & $-$33:26:22.50 & $1.57$  & $0.93$  & $0.322$ & A     & 2        & I        & 42\\
44     & B1600+434     & 16:01:40.45  & $+$43:16:47.8  & $1.59$  & $0.41$  & $0.690$ & A     & 2        & I        & 31\\
45     & B1608+656     & 16:09:13.96  & $+$65:32:29.0  & $1.39$  & $0.63$  & $1.135$ & A     & 4        & I        & 43\\
46     & FBQ1633+3134  & 16:33:48.99  & $+$31:34:11.90 & $1.52$  & $0.68$  & $0.330$ & B     & 2        & I        & 44\\
\hline
\end{tabular}
\caption{The columns are: (1) lens number; (2) lens name; (3)
R.A. [h:m:s], J2000.0; (4) Dec. [d:m:s], J2000.0; (5) source redshift,
$z_{s}$; (6) lens redshift, $z_{l}$; (7) critical radius, $r$, in
arcseconds; (8) grade for the likelihood that the object is a lens:
A=I'd bet my life, B=I'd bet your life, and C=I'd bet your life and
you should worry (CASTLES); (9) Number of images corresponding to each
source component, E means extended and R means there is an Einstein
ring (CASTLES); (10) sample: A1- Sample~A1 in this paper, C- Sample~C
in this paper, I- Sample~I in ORM, O- Targeted optical discoveries, R-
Targeted radio discoveries, M- Miscellaneous discoveries; (11)
references. The numbers shown are the ones actually used in the
computations. When computing ORM Sample I, we used the numbers from
ORM Table A1 (calculating the critical radius, $r$, as half of the
separation).
\newline
List of references:
\newline $1$ - Wisotzki et al. (2004); $2$ - Ofek et al. (2006); $3$ -
Rusin et al. (2003); $4$ - Morgan et al. (1999); $5$ - Wisotzki et
al. (2002); $6$ - Hagen \& Reimers (2000); $7$ - Johnston et
al.(2003); $8$ - Kneib, Cohen \& Hjorth (2000); $9$ - Kochanek et
al. (2000); $10$ - Lubin et al. (2000); $11$ - Eigenbrod et
al. (2006a); $12$ - Schechter et al. (1998); $13$ - Lehar et
al. (2000); $14$ - Lacy et al. (2002); $15$ - Hewett et al. (1994);
$16$ - Surdej et al. (1997); $17$ - Eisenhardt et al. (1996); $18$ -
Lidman et al. (2000); $19$ - Weymann et al. (1980); $20$ - Tonry
(1998); $21$ - Sluse et al. (2003); $22$ - Oguri et al. (2005); $23$ -
Eigenbrod et al. (2006b); $24$ - Burud et al. (2002a); $25$ - Wisotzki
et al. (1996); $26$ - Burud et al. (2002b); $27$ - Cohen, Lawrence \&
Blandford (2003); $28$ - Wiklind \& Combes (1995); $29$ - Lawrence et
al. (1995); $30$ - Tonry \& Kochanek (1999); $31$ - Fassnacht \& Cohen
(1998); $32$ - Myers et al. (1999); $33$ - Patnaik et al. (1992); $34$
- Lehar et al. (1993); $35$ - Treu \& Koopmans (2003); $36$ - Winn et
al. (2002); $37$ - Langston et al. (1989); $38$ - Wiklind \& Combes
(1996); $39$ - Sykes et al. (1998); $40$ - Warren et al. (1996); $41$
- Ratnatunga et al. (1998); $42$ - Gregg et al. (2000); $43$ -
Fassnacht et al. (1996); $44$ - Morgan et al. (2001); $45$ - Eigenbrod
et al. (2007); $46$ - Surdej et al. (1993); $47$ - Morgan et
al. (2005).}
\end{table*}

\newpage

\begin{table*}
\centering
\begin{tabular}{llllll}
\hline
Number & Name                       & $z_{s}$ & $z_{l}$ & $\sigma_a$ & Status\\
(1)    & (2)                        & (3)     & (4)     & (5) & (6)\\
\hline
1      & SDSS J003753.21$-$094220.1 & 0.632   & 0.195  & $265 \pm 10$ & C\\
2      & SDSS J021652.54$-$081345.3 & 0.524   & 0.332  & $332 \pm 23$ & C\\
3      & SDSS J073728.45$+$321618.5 & 0.581   & 0.322  & $310 \pm 15$ & C\\
4      & SDSS J081931.92$+$453444.8 & 0.446   & 0.194  & $231 \pm 16$ & UC\\
5      & SDSS J091205.30$+$002901.1 & 0.324   & 0.164  & $313 \pm 12$ & C\\
6      & SDSS J095320.42$+$520543.7 & 0.467   & 0.131  & $207 \pm 14$ & UC\\
7      & SDSS J095629.77$+$510006.6 & 0.470   & 0.241  & $299 \pm 16$ & C\\
8      & SDSS J095944.07$+$041017.0 & 0.535   & 0.126  & $212 \pm 12$ & C\\
9      & SDSS J102551.31$-$003517.4 & 0.276   & 0.159  & $247 \pm 11$ & C\\
10     & SDSS J111739.60$+$053413.9 & 0.823   & 0.229  & $279 \pm 21$ & UC\\
11     & SDSS J120540.43$+$491029.3 & 0.481   & 0.215  & $235 \pm 10$ & UC\\
12     & SDSS J125028.25$+$052349.0 & 0.795   & 0.232  & $254 \pm 14$ & C\\
13     & SDSS J125135.70$-$020805.1 & 0.784   & 0.224  & $216 \pm 23$ & C\\
14     & SDSS J125919.05$+$613408.6 & 0.449   & 0.233  & $263 \pm 17$ & UC\\
15     & SDSS J133045.53$-$014841.6 & 0.712   & 0.081  & $178 \pm 09$ & C\\
16     & SDSS J140228.21$+$632133.5 & 0.481   & 0.205  & $275 \pm 15$ & C\\
17     & SDSS J142015.85$+$601914.8 & 0.535   & 0.063  & $194 \pm 05$ & C\\
18     & SDSS J154731.22$+$572000.0 & 0.396   & 0.188  & $243 \pm 11$ & UC\\
19     & SDSS J161843.10$+$435327.4 & 0.666   & 0.199  & $257 \pm 25$ & C\\
20     & SDSS J162746.44$-$005357.5 & 0.524   & 0.208  & $275 \pm 12$ & C\\
21     & SDSS J163028.15$+$452036.2 & 0.793   & 0.248  & $260 \pm 16$ & C\\
22     & SDSS J163602.61$+$470729.5 & 0.675   & 0.228  & $221 \pm 15$ & UC\\
23     & SDSS J170216.76$+$332044.7 & 0.436   & 0.178  & $239 \pm 14$ & UC\\
24     & SDSS J171837.39$+$642452.2 & 0.737   & 0.090  & $270 \pm 16$ & C\\
25     & SDSS J230053.14$+$002237.9 & 0.464   & 0.229  & $283 \pm 18$ & C\\
26     & SDSS J230321.72$+$142217.9 & 0.517   & 0.155  & $260 \pm 15$ & C\\
27     & SDSS J232120.93$-$093910.2 & 0.532   & 0.082  & $236 \pm 07$ & C\\
28     & SDSS J234728.08$-$000521.2 & 0.715   & 0.417  & $330 \pm 50$ & UC\\
\hline
\end{tabular}
\caption{List of all lenses belonging to Bolton et al. (2006)
sample. The columns are: (1) lens number; (2) lens name; (3) source
redshift, $z_{s}$; (4) lens redshift, $z_{l}$; (5) velocity dispersion
measured within an aperture $\sigma_a$, in $\hbox{ km s}^{-1}$ and (6)
the status of the lens (C - confirmed; UC - unconfirmed). The numbers
shown are the ones actually used in the computations (only the central
value of $\sigma_a$ was used). The critical radius $r$ was then
computed using equation (2) and $h=0.7$, $\Omega_{\Lambda}=0.7$,
$\Omega_M=0.3$, and $\Omega_K=0.0$. Note that the 19 confirmed lenses
and 9 candidates are denoted with a $C$ and $UC$ respectively in the
final column.}
\end{table*}

\begin{figure*}
\begin{center}
\includegraphics[width=8.5cm]{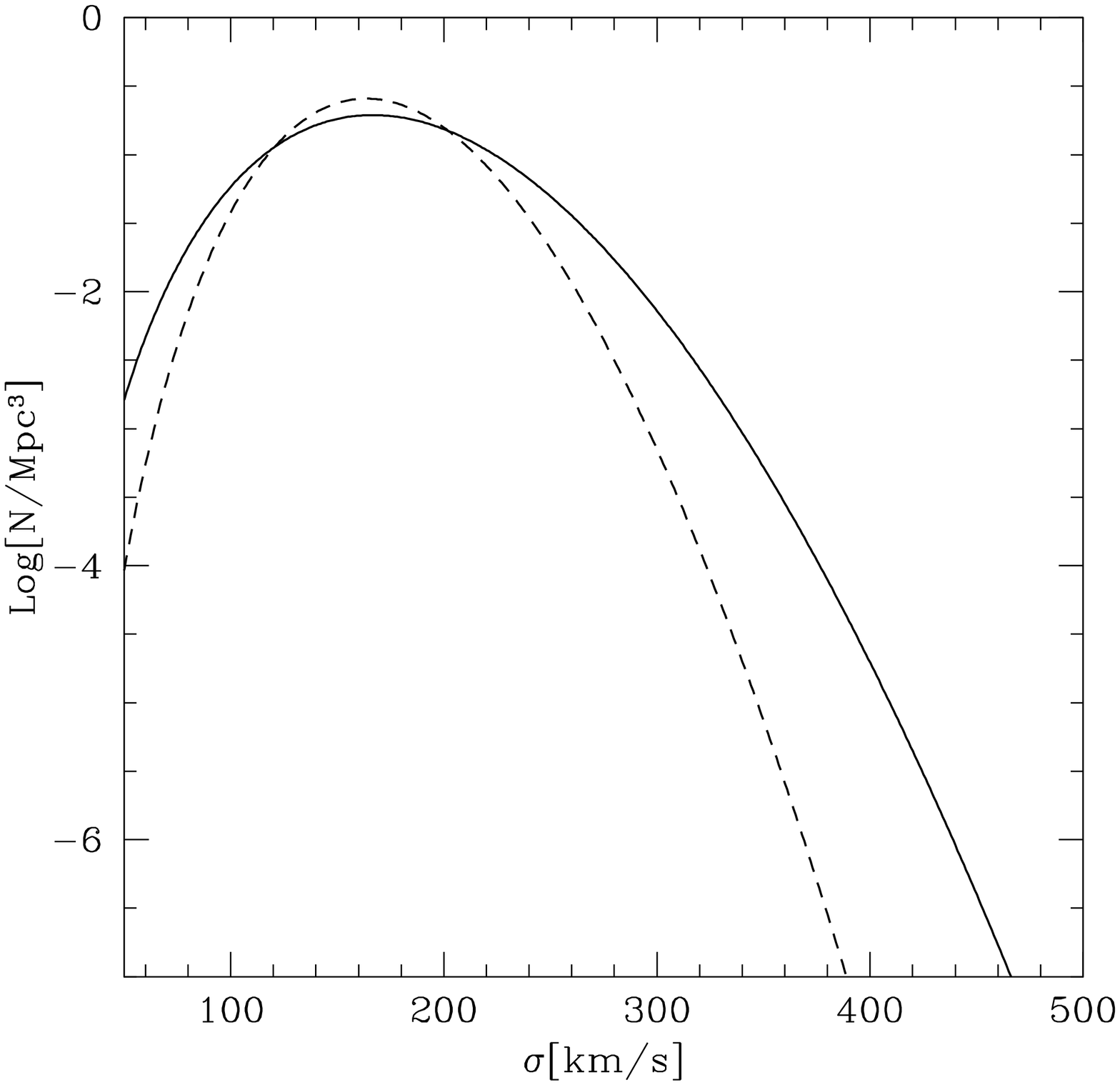}
\caption{The measured velocity dispersion function (MVDF) and the
inferred velocity dispersion function (IVDF) for early-type
galaxies. The functional forms plotted here are derived from fits
provided in equation (23) of Mitchell et al. (2005). The solid curve
is the MVDF and the dashed curve is the IVDF.}
\end{center}
\end{figure*}

\begin{figure*}
\begin{center}
\includegraphics[width=16cm,height=20cm]{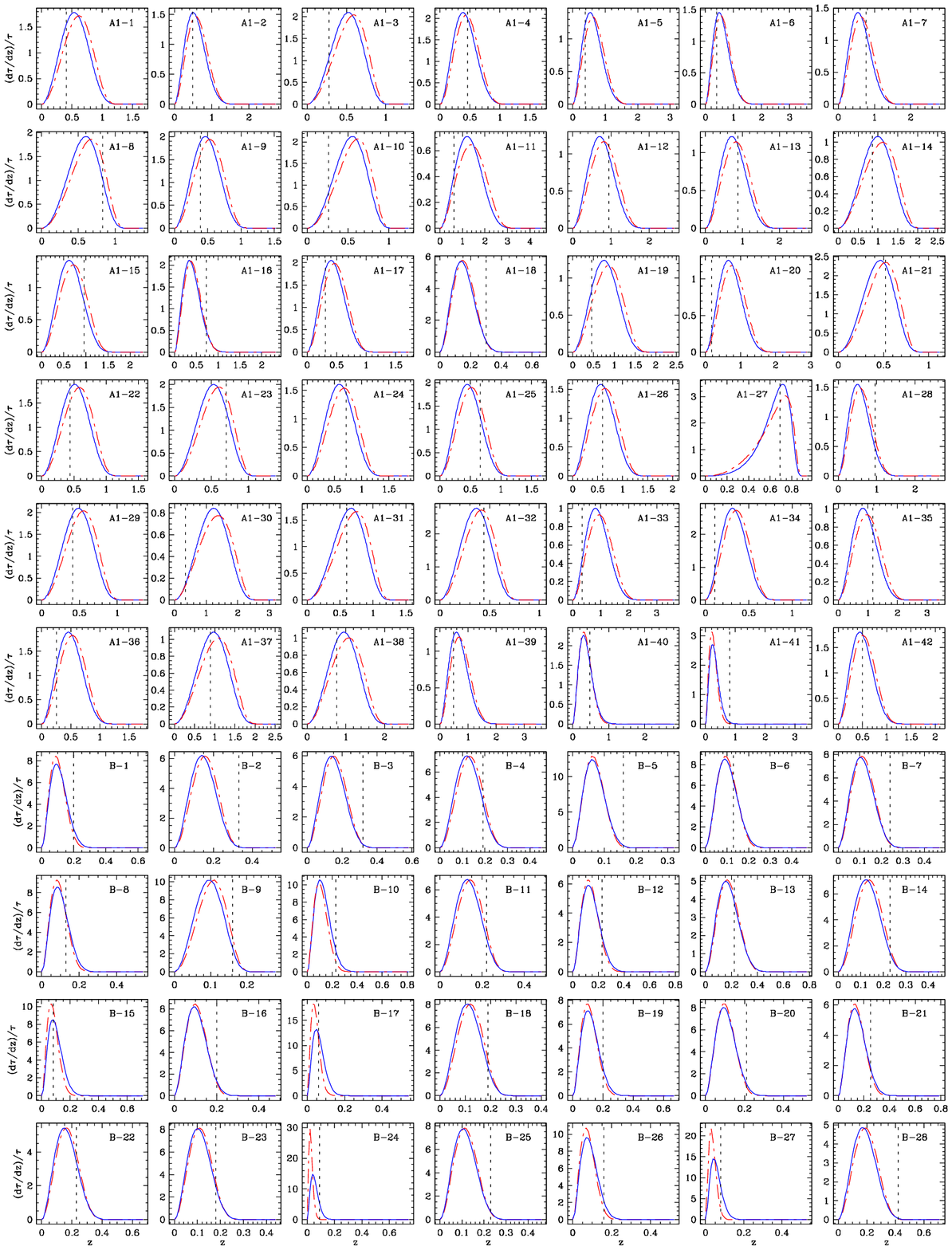}
\caption{The lens-redshift distribution (${(d\tau/dz)}/\tau$ vs $z$)
for all 70 lenses from Sample A1 (numbered in the same order as
presented in Table A1 in the Appendix) and the Bolton at al. (2006)
SLACS sample (numbered in the same order as presented in Table A2 in
the Appendix): calculated using the IVDF (red, dot-dashed line) and
the MVDF (blue, solid line); with the galaxy evolution and
cosmological parameters set to $U=P=Q=0$, $h=0.7$,
$\Omega_{\Lambda}=0.7$, $\Omega_M=0.3$, and $\Omega_K=0.0$. The
vertical dashed line marks the position of the observed lens redshift
$z_l$. The peak $z_p$ of the probability distribution is skewed to
slightly higher $z_l$ for the IVDF compared to the MVDF, in most
lenses.}
\end{center}
\end{figure*}
\label{lastpage}

\end{document}